\newcommand\ifApJ[2]{#1}  
\renewcommand\ifApJ[2]{#2}  

\documentclass[apj,iop,12pt]{emulateapj}
\usepackage{apjfonts}
\usepackage{amsmath,amsbsy}
\usepackage{color}
\definecolor{darkblue}{rgb}{0,0,0.5}
\usepackage[pdftitle={Chandra X-ray spectroscopy of the focused wind in the Cygnus X-1 system},
	    pdfsubject={I. The nondip spectrum in the low/hard state},
	    pdfauthor={Manfred Hanke et al.},
	    pdfkeywords={accretion, accretion disks -- stars: individual (HDE 226868, Cyg X-1) -- stars: winds, outflows -- techniques: spectroscopic -- X-rays: binaries},
            colorlinks, linkcolor=black, urlcolor=darkblue, citecolor=black
	   ]{hyperref}
\renewcommand{\url}[1]{\href{#1}{#1}}
\newcommand\Section{Section}
\newcommand\Figure{Figure}
\newcommand\Equation{Equation}
\newcommand\Sect[1]{\Section~\ref{#1}}
\newcommand\Fig[1]{\Figure~\ref{#1}}
\newcommand\Eq[1]{\Equation~(\ref{#1})}
\newcommand\Eqs[1]{Equations~(\ref{#1})}

\bibpunct{(}{)}{;}{a}{}{,}
\newcommand\Cyg{\mbox{Cyg\,X-1}}
\newcommand\RXTE{\textsl{RXTE}}
\newcommand\Chandra{\textsl{Chandra}}
\newcommand\Ion[2]{#1\,\textsc{#2}}
\newcommand\eVS{\raisebox{-1.2ex}{\rule{0pt}{4ex}}}  
\newcommand\boldvec[1]{\pmb{#1}}  
\newcommand\vecv{\boldvec{v}}
\newcommand\vecr{\boldvec{r}}

\begin{document}
\journalinfo{The Astrophysical Journal, \rm 690:330--346, 2009 January 1 \hfill
 \href{http://dx.doi.org/10.1088/0004-637X/690/1/330}{doi:10.1088/0004-637X/690/1/330}
}
\title{~\\[-1.2cm]\large
 \textsl{Chandra} X-ray spectroscopy of the focused wind in the Cygnus~X-1 system\\
 I. The nondip spectrum in the low/hard state
 \vskip4mm
}
\shorttitle{\scshape \textsl{CHANDRA} X-RAY SPECTROSCOPY OF THE FOCUSED WIND IN THE CYG~X-1 SYSTEM. \;\; I.}
\shortauthors{\scshape HANKE ET AL.}
\author{Manfred~Hanke\altaffilmark{1,2}, J\"orn~Wilms\altaffilmark{1,2},
        Michael~A.~Nowak\altaffilmark{3},
        Katja~Pottschmidt\altaffilmark{4,5,6},
        Norbert~S.~Schulz\altaffilmark{3},
        and Julia~C.~Lee\altaffilmark{7}
}
\email{Manfred.Hanke@sternwarte.uni-erlangen.de}
\altaffiltext{1}{Dr.~Karl Remeis-Sternwarte, Astronomisches Institut der Universit\"at Erlangen-N\"urnberg, Sternwartstr.~7, 96049 Bamberg, Germany}
\altaffiltext{2}{Erlangen Centre for Astroparticle Physics, University of Erlangen-Nuremberg, Erwin-Rommel-Stra\ss e 1, 91058 Erlangen, Germany}
\altaffiltext{3}{MIT-CXC, NE80-6077, 77 Mass. Ave., Cambridge, MA 02139, USA}
\altaffiltext{4}{CRESST, University of Maryland Baltimore County, 1000 Hilltop Circle, Baltimore, MD 21250, USA}
\altaffiltext{5}{NASA Goddard Space Flight Center, Astrophysics Science Division, Code 661, Greenbelt, MD 20771, USA} 
\altaffiltext{6}{Center for Astrophysics and Space Sciences, University of California at San Diego, La Jolla, 9500 Gilman Drive, CA 92093-0424, USA}
\altaffiltext{7}{Harvard University, Department of Astronomy (part of the Harvard-Smithsonian Center for Astrophysics), 60 Garden Street, MS-6, Cambridge, MA 02138, USA}
\submitted{\scriptsize Received 2008 May 28; accepted 2008 August 27; published 2008 December 1}

\begin{abstract}\noindent
  We present analyses of a 50\,ks observation of the supergiant X-ray binary system
  \ifApJ{Cygnus\,X-1 (\mbox{Cyg\,X-1})/\linebreak}{Cygnus\,X-1/}HDE\,226868
  taken~with the \Chandra{} High Energy Transmission Grating Spectrometer (HETGS).
  Cyg\,X-1 was in its spectrally hard state and
  the observation was performed during superior conjunction of the black hole,
  allowing for the spectroscopic analysis of the accreted stellar wind along the line of sight.
  A significant part of the observation covers X-ray dips as commonly observed for \Cyg{}
  at this orbital phase, however, here we analyze only the high count rate nondip spectrum.
  The full 0.5--10\,keV continuum can be described by a single model 
  consisting of a disk, a narrow and a relativistically broadened Fe K$\alpha$ line,
  and a power-law component,
  which is consistent with simultaneous 
  \ifApJ{\textsl{Rossi X-Ray Timing Explorer}}{\RXTE} broad band data.
  We detect absorption edges from overabundant neutral O, Ne, and Fe,
  and absorption line series from highly ionized ions and infer column densities and Doppler shifts.
  With emission lines of He-like Mg\,\textsc{xi}, we detect two plasma components
  with velocities and densities consistent with the base of the spherical wind and a focused wind.
  A simple simulation of the photoionization zone suggests
  that large parts of the spherical wind outside of the focused stream are completely ionized,
  which is consistent with the low velocities ($<$200\,km\,s$^{-1}$) observed in the absorption lines,
  as the position of absorbers in a spherical wind at low projected velocity is well constrained.
  Our observations provide input for models that couple the wind activity of HDE\,226868
  to the properties of the accretion flow onto the black hole.
\end{abstract}
\keywords{
   accretion, accretion disks
-- stars: individual (HDE\,226868, \mbox{Cyg\,X-1})
-- stars: winds, outflows
-- techniques: spectroscopic
-- X-rays: binaries
\ifApJ{\\\textsl{Online-only material:} color figures}{}
}

\section{Introduction} \label{sec:intro}
\object[Cygnus X-1]{Cygnus\,X-1\ifApJ{ (Cyg X-1)}{}} was discovered in 1964 \citep{Bowyer1965}
and soon identified as a high-mass X-ray binary system (HMXB)
with an orbital period of 5.6\,d
\citep{MurdinWebster1971,WebsterMurdin1972,Bolton1972}.
It consists of the supergiant O9.7 star \object[HDE 226868]{HDE\,226868}
\citep{Walborn1973,Humphreys1978}
and a compact object, which is dynamically constrained
to be a black hole \citep{GiesBolton1982}.
The detailed spectroscopic analysis of HDE\,226868 by \citet{Herrero1995}
gives a stellar mass $M_\star\approx18\,M_\odot$,
leading to a mass of $M_\mathrm{BH}\sim$10\,$M_\odot$ for the black hole,
if an inclination $i\approx35^\circ$ is assumed.
Note that \citet{Ziolkowski2005} derives a mass of $M_\star=(40\pm5)\,M_\odot$
from the evolutionary state of HDE\,226868, corresponding to $M_\mathrm{BH}=(20\pm5)\,M_\odot$,
while \citet{Shaposhnikov2007} claim $M_\mathrm{BH}=(8.7\pm0.8)\,M_\odot$
from X-ray spectral-timing relations.

\Cyg{} is usually found in one of the two states
that are distinguished by the soft X-ray luminosity and spectral shape,
the timing properties, and the radio flux
\citep[see, e.g.,][]{Pottschmidt2003,Gleissner2004_III,Gleissner2004_II,Wilms2006}:
the low/hard state is characterized by a lower luminosity below 10\,keV,
a hard Comptonization power-law spectrum (photon index $\Gamma\sim1.7$)
with a cutoff at high energies (folding energy $E_\mathrm{fold}\sim150\,$keV)
and strong variability of $\sim$30\% root mean square (rms).
Radio emission is detected at the $\sim$15\,mJy level.
In the high/soft state, 
the soft X-ray spectrum is dominated by a bright and much less variable
(only few~\% rms) thermal disk component,
and the source is invisible in the radio.
Within the classification of \citet{RemillardMcClintock2006},
the high/soft state of \Cyg{} corresponds to the steep power-law state
rather than to the thermal state,
as a power-law spectrum with photon index $\Gamma\sim2.5$
may extend up to $\sim$10\,MeV \citep{Zhang1997,McConnell2002,CadolleBel2006}.
Most of the time, \Cyg{} is found in the hard state,
but transitions to the soft state and back after a few weeks or months
are common every few years.
Transitional or intermediate states \citep{Belloni1996}
are often accompanied by radio and/or X-ray flares.
Similar to a transition to the soft state,
the spectrum softens during these flares and the variability is reduced.
This behavior is called a ``failed state transition''
if the true soft state is not reached
\citep{Pottschmidt2000,Pottschmidt2003}.
Transitional states have occurred more frequently since mid-1999
than before \citep{Wilms2006},
which might indicate changes in the mass-accretion rate
due to a slight expansion of HDE\,226868 \citep{Karitskaya2006}.

HMXBs are believed to be powered by accretion from the stellar wind.
The accretion rate and therefore X-ray luminosity and spectral state
are thus very sensitive to the wind's detailed properties
such as velocity, density, and ionization.
For HDE\,226868, \citet{Gies2003} found an anticorrelation
between the H$\alpha$ equivalent width
(an indicator for the wind mass loss rate $\dot M_\star$)
and the X-ray flux.
Considering the photoionization of the wind
would allow for a self-consistent explanation
\citep[see, e.g.,][]{Blondin1994}: a \emph{lower} mass loss gives
a \emph{lower} wind density and therefore \emph{higher} degree of ionization
due to the irradiation of hard X-rays,
which reduces the driving force of HDE\,226868's UV photons on the wind
and results in a \emph{lower} wind velocity $v$,
leading finally to a \emph{higher} accretion rate
\citep[$\propto\dot M_\star/v^4$,][]{BondiHoyle1944}.
However, \citet{Gies2008} find suggestions
that the photoionization and velocity of the wind
might be similar during both hard and soft states.
UV observations allow the photoionization
in the HDE\,226868\,/\,\Cyg{} system to be probed:
\citet{Vrtilek2008} reported P\,Cygni profiles
of \Ion{N}{v}, \Ion{C}{iv}, and \Ion{Si}{iv}
with weaker absorption components at orbital phase $\phi_\mathrm{orb}\approx0.5$,
i.e., when the black hole is in the foreground of the supergiant.
This reduced absorption, which was already found by \citet{Treves1980},
is due to the \citet{HatchettMcCray1977} effect,
showing that those ions become superionized by the X-ray source.
\citet{Gies2008} model the orbital variations of the UV lines
assuming that the wind of HDE\,226868 is restricted to the shadow wind
from the shielded side of the stellar surface \citep{Blondin1994},
i.e., the \citet{Stromgren1939} zone of \Cyg{} extends to the donor star.
However, this assumption applies only to the spherical part of the wind,
which might therefore hardly contribute to the mass accretion of \Cyg{}.

As HDE\,226868 is close to filling its 
Roche lobe \citep{Conti1978,GiesBolton1986_II,GiesBolton1986_III},
the wind is not spherically symmetrical as for isolated stars,
but strongly enhanced toward the black hole
\citep[``focused wind\ifApJ{;''}{'';}][]{FriendCastor1982}.
The strongest wind absorption lines in the optical 
are therefore observed at the conjunction phases \citep{Gies2003}.
Similarly, X-ray absorption dips occur preferentially around $\phi_\mathrm{orb}=0$,
i.e., during superior conjunction of the black hole \citep{BalucinskaChurch2000}.
These dips are probably caused by dense, neutral clumps,
formed in the focused wind where the photoionization is reduced,
although recent analyses have also suggested
that part of the dipping activity may result from the interaction of the focused wind
with the edge of the accretion disk \citep{Poutanen2008}.

The photoionization and dynamics of both the spherical and focused winds
can also be investigated with the high-resolution grating spectrometers
of the modern X-ray observatories \Chandra{} or \textsl{XMM-Newton}.
As none of the previously reported observations of \Cyg{}
was performed at orbital phase $\phi_\mathrm{orb}=0$ and in the hard state,
when the wind is probably denser and less ionized than in the soft state,
the \Chandra{} observation presented here allows for the most detailed
investigation of the focused wind to date.

The remainder of this paper is organized as follows:
in \Sect{sec:obs}, we describe our observations of \Cyg{}
with \Chandra{} and the \textsl{Rossi X-Ray Timing Explorer}
(\RXTE), and how we model CCD pile-up for the \Chandra-%
\ifApJ{High Energy Transmission Grating Spectrometer (HETGS)}{HETGS} data.
We present our investigations in \Sect{sec:analysis}:
after investigating the light curves, we model the nondip continuum
and analyze neutral absorption edges and
absorption lines from the highly ionized stellar wind
-- and the few emission lines from He-like ions, which indicate two plasma components.
In \Sect{sec:discuss}, we discuss models for the stellar wind and the photoionization zone.
We summarize our results
after comparing them with those of the previous \Chandra{} observations of \Cyg.

\section{Observation and Data Reduction} \label{sec:obs}
\begin{table}{\centering
 \caption{Observations of \Cyg{}}
 \label{tab:observations}
 \begin{tabular}{cccccc}
  \hline
  \hline
  Satellite /      & Start    & Stop     & Exposure\footnote{~For the \Chandra{} data, the nondip GTIs (see \Fig{fig:lc_hardness}) have been used.\ifApJ{ }{\\\indent~}For \RXTE{}, the 11th orbit was considered.}         & Count Rate$^a$ \\
  Instrument       & (MJD)    & (MJD)    & (ks)             & (cps)      \\
  \hline
  \Chandra{} & 52748.70 & 52749.28 & (47.2)& (\nodata) \\
  MEG$\pm$1\footnote{~The \Chandra-MEG spectra cover $\approx$\,0.8--6\,keV (1--99\,\% quantiles).} &(52748.70)&(52749.28)&  16.1 & $2\times27$\\
  HEG$\pm$1\footnote{~The \Chandra-HEG spectra cover $\approx$\,1--7.5\,keV (1--99\,\% quantiles).} &(52748.70)&(52749.28)&  16.1 & $2\times17$ \\
  \hline
  \RXTE{}    & 52748.08 & 52749.18 & \nodata & \nodata \\
  PCA\footnote{~The \RXTE-PCA data from 4 to 20\,keV has been used.}        & 52748.74 & 52748.78 &  3.0 & 1456 \\
  HEXTE a+b\footnote{~The \RXTE-HEXTE data from 20 to 250\,keV has been used.}  & 52748.74 & 52748.78 &  1.1 & 2$\times186$ \\
  \hline
 \end{tabular}\\
 \ \\
 }{\bfseries Notes.}
\end{table}
\begin{figure}\centering
 \includegraphics[width=0.95\columnwidth]{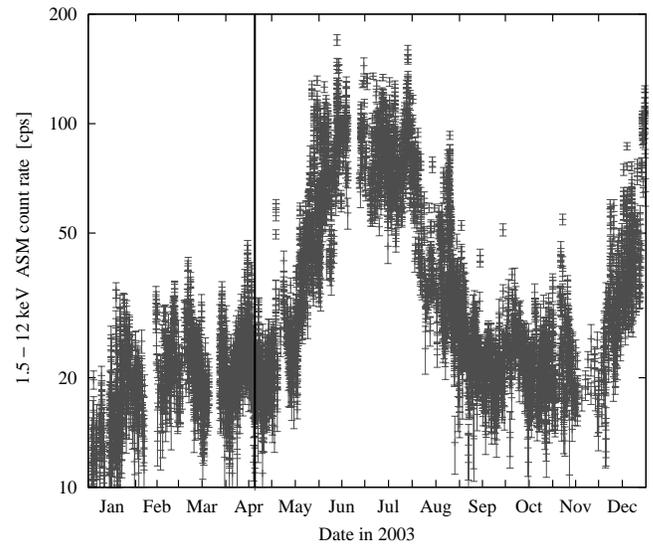}
 \caption{Brightness of \Cyg{} as seen by the ASM on board \RXTE{}.
          During the \Chandra{} observation (marked by a line),
          the source was still in its low/hard state:
          the 1.5--12\,keV count rate did not exceed 25\,cps.
          In spite of the high intrinsic variability, the high/soft state 
          during June, July, and August can clearly be distinguished,
          a result which is also found by spectral analysis \citep{Wilms2006}.
         }
 \label{fig:ASM}
\end{figure}

\subsection{\Chandra{} ACIS-S/HETGS Observation} \label{sec:ChandraObs}
\Cyg{} was observed on 2003 April 19 and 20
by the \Chandra{} X-Ray Observatory, see Table~\ref{tab:observations}.
An overview on all its instruments is given by the Proposers' Observatory Guide \citep{CXC_POG}.
In the first four months of 2003, the \RXTE{} All-Sky Monitor (ASM;
see \citealt{Doty1994,Levine1996}) 
showed the source's 1.5--12\,keV count rate to be generally below 50\,ASM-cps (\Fig{fig:ASM}).
At the time of our \Chandra{} observation, it was less than 25\,cps (0.33\,Crab),
typically
indicative of the source being in its low/hard state \citep{Wilms2006}.

\ifApJ{HETGS}{The High Energy Transmission Grating Spectrometer (HETGS),}
containing high and medium energy gratings (HEG/MEG; see \citealp{Canizares2005})
was used to disperse X-ray spectra with the highest resolution \citep[Table 8.1]{CXC_POG}:
\begin{equation}
 \Delta\lambda_\mathrm{HEG} \;\;=\;\; 5.5\,\mbox{m\AA \quad and\quad} \Delta\lambda_\mathrm{MEG} = 11\,\mbox{m\AA}
 \label{eq:deltaLambda}
\end{equation}
As only half of the spectroscopy array of the
Advanced CCD Imaging Spectrometer \citep[ACIS-S; see][]{Garmire2003},
namely a 512 pixel broad subarray,
was operated in the timed event (TE) mode,
the six CCDs could be read out after exposure times of $t_\mathrm{frame}=1.7\,$s
and the position of each event is well determined.
Photons from the different gratings can thus easily be distinguished
due to the different dispersion directions of the HEG and the MEG.
Even in its low state, however, \Cyg{} is so bright
that several photons may pile up in a CCD pixel during one readout frame.
Both events cannot be discriminated and are interpreted as a single photon
with larger energy.
As the undispersed image would have been completely piled up,
only 10\% of those events in a $40\times38$ pixel window have been transmitted
in order to save telemetry capacity.
The first-order spectra are, however, only moderately affected,
which can be modeled very well (\Sect{sec:pile-up}).
The alternative to this approach would have been to use the continuous clocking (CC) mode,
where the ACIS chips are read out continuously in 2.85\,ms,
but only the position 
perpendicular to the readout direction
can be determined for every photon event.
The CC mode was, however, avoided due to difficulties
in the reconstruction of HEG and MEG spectra
and other calibration issues.

The undispersed position of the source is required for the wavelength calibration
of the spectra. We redetermined it to $\mathrm{R.A.}=19^\mathrm{h}\,58^\mathrm{m}\,21\fs67$,
$\delta=+35\degr\,12\arcmin\,5\farcs83$
from the intersection of the HEG and MEG arm and the readout streak
\citep{Bish2006}.
Afterward, the event lists were reduced using the standard software from the 
\Chandra{} X-ray Center (CXC), \texttt{CIAO}\,3.3 with \texttt{CALDB}\,3.2.3.\footnote{%
 See \url{http://cxc.harvard.edu/ciao3.3/}\ifApJ{}{.}}
Exceptionally narrow extraction regions had to be chosen
as the background spectrum would otherwise have been dominated
by the dispersed extended X-ray-scattering halo around the source
\citep{Xiang2005}.
The further analysis was performed with the Interactive Spectral Interpretation System
(\textsc{ISIS})~1.4.9 \citep{Houck2002}.\footnote{%
 See \url{http://space.mit.edu/cxc/isis/}\ifApJ{}{.}}

We use the four first-order MEG and HEG spectra
(with two dispersion directions each, called $+1$ and $-1$ in the following)
which provide the best signal-to-noise ratio (S/N).
The ``second-order spectra'' are dominantly formed
by piled first-order events which reach the other order sorting window 
of data extraction (defined in dispersion-energy space)
when the energy of two first-order photons accumulates.
This effect is most evident for the MEG,
whose even dispersion orders are suppressed
by construction of the grating bars 
\citep{CXC_POG}.

\subsection{Model for Pile-Up in Grating Observations} \label{sec:pile-up}
For the first-order spectra, pile-up causes a pure reduction of count rate:
a multiple event,
i.e., the detection of more than one photon in a CCD pixel during one readout time,
which cannot be separated,
is either rejected by grade selection during the data processing
or migrates to a higher-order spectrum.
The Poisson probability for single events
in a $3\times3$ pixel event-detection cell $i$
\citep[see][]{Davis2002,Davis2003,CXC2005_pileupABC}
is
\begin{equation}
 P_1(i) \;\;=\;\; \Lambda(i) \:\cdot\: \exp\big(-\Lambda(i)\big) \;\;,
 \label{eq:P1}
\end{equation}
where the expected number of events,
$\Lambda(i) = \gamma_0\cdot C_\mathrm{tot}(i)$,
is given by the total spectral count rate, $C_\mathrm{tot}(i)$,
at this position (in units of counts per \AA{} and s),
and where the constant $\gamma_0$ is
\begin{equation}
 \gamma_0 \;\;=\;\; 3\;\Delta\lambda \cdot t_\mathrm{frame} \;\;,
 \label{eq:pile-up-scale}
\end{equation}
where $\Delta\lambda$ is the resolution of the spectrometer of \Eq{eq:deltaLambda},
and $t_\mathrm{frame}$ is the frame time.
We therefore describe the pile-up in the first-order spectra
with the nonlinear convolution model \texttt{simple\_gpile2} in \textsc{ISIS},\footnote{
 The \textsc{ISIS}/\textsc{S-Lang} code for \texttt{simple\_gpile2} is available online at
 \url{http://pulsar.sternwarte.uni-erlangen.de/hanke/X-ray/code/simple\_gpile2.sl}\ifApJ{}{.}}
which exponentially reduces the predicted count rate $C(\lambda)$
according to \Eq{eq:P1}:
\begin{equation}
  C'(\lambda) \;\;=\;\; C(\lambda) \:\cdot\: \exp\big(-\gamma\cdot C_\mathrm{tot}(\lambda)\big)
 \label{eq:pile-up}
\end{equation}
Here, the scale $\gamma\approx\gamma_0$ is left as fit parameter,
and $C_\mathrm{tot}(\lambda)$ also takes the photons into account
which are dispersed in a higher order $m\le3$.
The count rates are estimated from the corresponding effective areas $A_m$
and the assumed photon flux $S$:
\begin{equation}
 C_\mathrm{tot}(\lambda) \;\;=\;\; \sum\nolimits_{m=1}^3 A_m(\lambda/m) \:\cdot\: S(\lambda/m)
 \label{eq:pile-up_estimate}
\end{equation}
\texttt{simple\_gpile2} is based on the \texttt{simple\_gpile} model
\citep{CXC2005_pileupABC,Nowak2008}, which parameterizes the strength of pile-up
by the (maximum) pile-up fraction \mbox{$p = 1-\exp\left(-\gamma\cdot \max\{C_\mathrm{tot}\}\right)$},
while using the parameter $\gamma$ of \texttt{simple\_gpile2} 
avoids to have a nonlocal model which depends on the flux at the position of the highest pile-up.
\begin{figure}\centering
 \includegraphics[width=0.95\columnwidth]{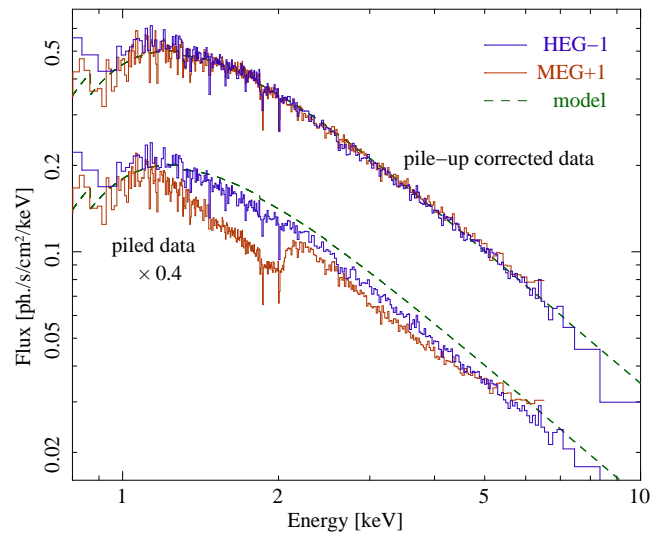}
 \caption{Pile-up in the HEG (black curves) and MEG (gray curves) data:
          the lower spectra, which are shifted by a factor of 0.4,
          show the uncorrected data. The dashed line shows the model (free of pile-up).
          The MEG spectrum suffers from significant 
          pile-up losses around 2\,keV, where the highest count rate is obtained.
          The upper spectra show the pile-up corrected spectra.
          Note that we show \textsc{ISIS}' (model independent) flux spectra only for illustration;
          \texttt{simple\_gpile}(\texttt{2}) operates on the count rates predicted by a model.
         }
 \label{fig:fluxspectrum}
\end{figure}

The effect of pile-up is stronger in the MEG spectra than in the HEG spectra
due to the lower dispersion and higher effective area of the MEG.
The apparent flux reduction is most significant near 2\,keV where the spectrometer has the largest efficiency
and the highest count rates are obtained (see \Fig{fig:fluxspectrum}).
It can, however, excellently be modeled with \verb|simple_gpile2|.
When fitting a spectrum, there is always a strong correlation between the pile-up scale $\gamma$
and the corresponding flux normalization factor,
e.g., the relative cross-calibration factor $c$ introduced in \Sect{sec:continuum}.
The best-fit values for the pile-up scales $\gamma$
found in our data analysis (see Table~\ref{tab:continuum})
are only slightly larger than expected from \Eq{eq:pile-up-scale}
(namely $\gamma_{0, {\rm MEG}\,\pm1} = 5.6\!\times\!10^{-2}\:$s$\;$\AA{}
and $\gamma_{0, {\rm HEG}\,\pm1} = 2.8\!\times\!10^{-2}\:$s$\;$\AA),
and the calibration factors $c$ are consistent with 1,
except for the HEG$-1$ spectrum,
for which both the largest $\gamma-\gamma_0$ and $c$ were found.
Given the presence of the $\gamma$--$c$ correlation,
we consider the latter to be a numerical artifact.

According to the \verb|simple_gpile2| model,
the MEG+1 spectra suffer from \ifApJ{greater than }{$>$}30\,\% pile-up for $6\,\mbox{\AA}\le\lambda\le8.3\,$\AA,
peaking at $p_\mathrm{MEG+1}=45\,\%$ in the Si\,\textsc{xiii} f emission line at 6.74\,\AA.
Except for some emission lines, among them the Fe\,K$\alpha$ line, 
the continuum pile-up fraction of the HEG spectra is below 17\,\%.
For the HEG+1 spectrum, the reduction is less than 10\,\%
outside the ranges $2.09\,\mbox{\AA}\le\lambda\le4.08\,$\AA{}
and $6.05\,\mbox{\AA}\le\lambda\le6.93\,$\AA{}.
\begin{figure*}\centering
 \includegraphics[width=0.95\textwidth]{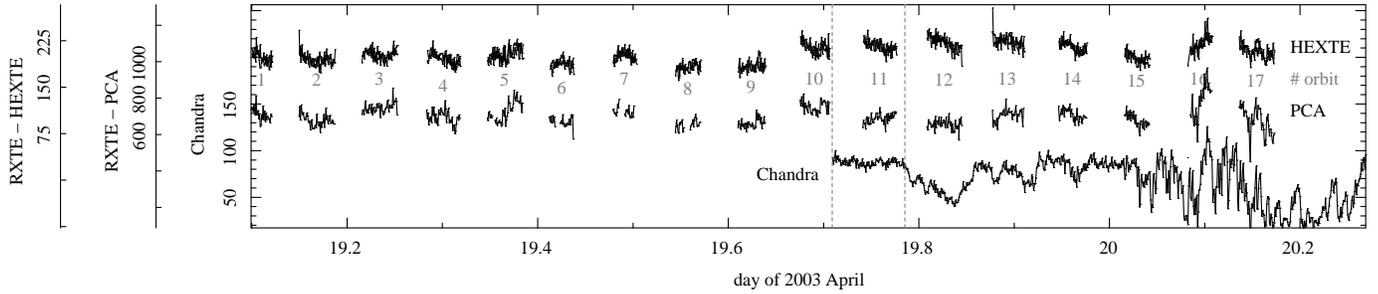}
 \caption{Coverage of the simultaneous \RXTE{} and \Chandra{} observations.
          The plot shows the background-subtracted light curves with a time resolution of 64\,s
          on a separate $y$ axis each.
          Top: \RXTE-HEXTE (20--250\,keV). The count rates have been corrected for the detector dead time.
          Center: \RXTE-PCA (4--20\,keV), normalized by the number of active PCUs.
          Bottom: \Chandra-HETGS (0.5--12\,keV, only first-order events;
                  see \Fig{fig:lc_hardness} for more details).
          The numbers label the \RXTE{} orbits,
          vertical lines mark the first part of the nondip spectrum
          (see \Sect{sec:lightcurve} and \Fig{fig:lc_hardness}),
          which completely covers \RXTE{} orbit 11.
         }
 \label{fig:RXTElightcurve}
\end{figure*}

\subsection{\RXTE{} Observation}
While \Chandra's HETGS provides a high spectral resolution,
the energy range covered is rather limited.
Within the framework of our \RXTE{} monitoring campaign,
the broadband spectrum of \Cyg{}
was measured regularly, i.e., at least biweekly, since 1998
\citep[see][and references therein]{Wilms2006}.
The observation on 2004 April 19 was extended
to provide hard X-ray data simultaneously with the \Chandra{} observation.
More than one day was covered by 17 \RXTE{} orbits of $\sim$47\,min good time,
interrupted by $\sim$49\,min intervals when \Cyg{} was not observable
due to Earth occultations or passages through the South Atlantic Anomaly (SAA),
see \Fig{fig:RXTElightcurve}.

Data in the 4--20\,keV range from the Proportional
Counter Array \citep[PCA;][]{Jahoda1996} and in the 20--250\,keV range
from the High Energy X-Ray Timing Experiment
\citep[HEXTE;][]{Gruber1996} were used. The data were extracted using
\textsc{HEASOFT}~6.3.1,\footnote{See \url{http://heasarc.gsfc.nasa.gov/lheasoft/}\ifApJ{}{.}}
following standard data screening procedures
as recommended by the \RXTE{} Guest Observer Facility.
Data were only used if taken more than 30\,minutes away from the SAA.
For the PCA, only data taken in the top layer of the proportional counter
were included in the final spectrum, and no additional systematic
error was added to the spectrum.
During the observation, 
different sets of Proportional Counter Units (PCUs) were operative.
During the 11th orbit, extensively used in this work,
PCUs 1 and 4 were off.

\section{Analysis} \label{sec:analysis}
\ifApJ{}{\enlargethispage{\baselineskip}}
\subsection{Light Curve} \label{sec:lightcurve}
\begin{figure}\centering
 \includegraphics[width=0.95\columnwidth]{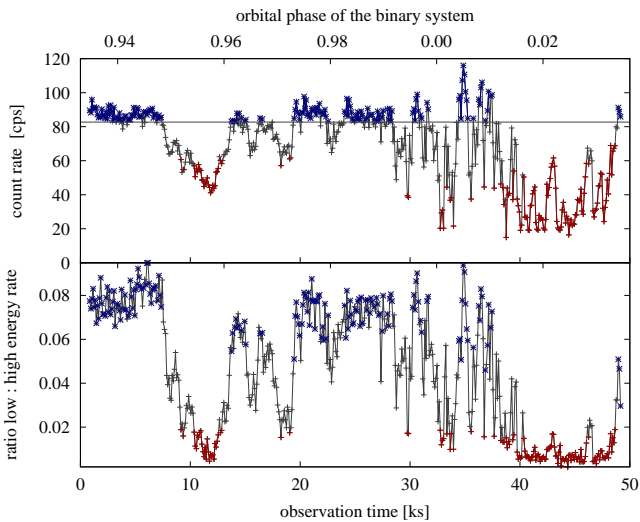}
 \caption{Top: full 0.5--12\,keV band \Chandra{} light curve.
          Bottom: ratio of 0.7--1\,keV band and 2.1--7.2\,keV band count rates.
          Absorption dips\ifApJ{---}{ -- }at first compact, then with complex substructure\ifApJ{---}{ -- }
          show up with a reduced flux and spectral hardening.
          Count~rates \ifApJ{greater than }{$>$}82.7\,cps define the nondip data (dark).
         }
 \label{fig:lc_hardness}
\end{figure}
The \Chandra{} observation covers a phase range between $\sim$0.93 and $\sim$0.03
in the 5.599829\,d binary orbit \citep[whose epoch is HJD\,$2451730.449\pm0.008$]{Gies2003}.
The top panel of \Fig{fig:lc_hardness} shows the light curve of first-order events (MEG$\pm$1, HEG$\pm1$)
in the full band accessible with \Chandra-HETGS.
During several dip events, the flux is considerably reduced.
The absorption dips distinguish themselves also by spectral hardening
(the bottom panel of \Fig{fig:lc_hardness}).
We extract a 16.1\,ks nondip spectrum, which is the subject of this paper,
from all times when the total count rate exceeds 82.7\,cps;
these are indicated by dark points in \Fig{fig:lc_hardness}.
The analysis of dip spectra will be described in a subsequent paper.

The \RXTE{} light curve shows considerable variability as well.
Although the dips are more obviously detected with \Chandra{}
in the soft X-ray band, similar structures are also seen with \RXTE-PCA or even -HEXTE,
especially in the last \RXTE{} orbits of these observations
(\Fig{fig:RXTElightcurve}).
We chose to infer the nondip broadband spectrum from the \RXTE{} data
taken during the 11th orbit,
which was performed entirely during the first part of the nondip phase.
Other parts are interrupted by dips, occultations, or have nonuniform PCU configuration.

\subsection{Continuum Spectrum} \label{sec:continuum}
\begin{figure}\centering
 \includegraphics[width=0.95\columnwidth]{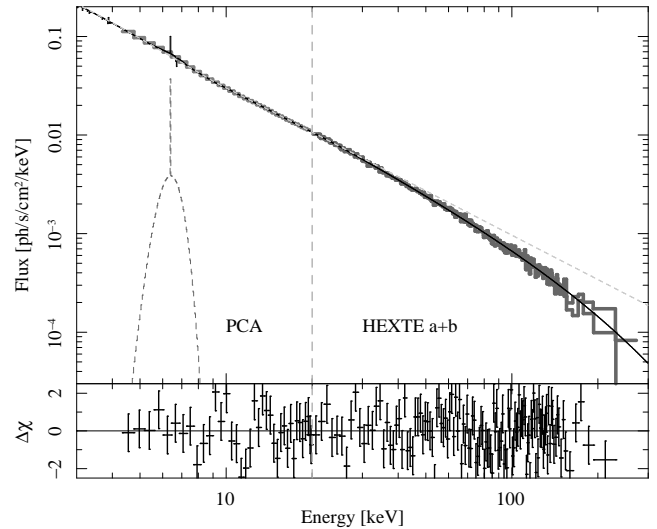}
 \caption{\RXTE{} 4--250\,keV broadband continuum spectrum,
          as measured with PCA below 20\,keV and above 20\,keV with HEXTE,
          which can be described by a broken power-law with high-energy cutoff
          and a weak iron line, see Table~\ref{tab:continuum}.
          Due to the joint fit with \Chandra, the Fe line consists
          of narrow and a broad component, see \Fig{fig:IronLine}.
          The HEXTE spectra shown are renormalized to \mbox{match the} PCA flux,
          as the absolute calibrations of these two instruments differ by $\sim$15\%.
         }
 \label{fig:RXTEspectrum}
\end{figure}
\begin{table}{\centering
 \caption{Fit Parameters of the Continuum Nondip Hard State Spectrum of \Cyg{}}
 \label{tab:continuum}
 \begin{tabular}{cccc}
  \hline
  \hline
   &  & Fit to the & Joint Fit to Both the \\
  Parameter & Unit & \textsl{Chandra} & \textsl{Chandra} and \textsl{RXTE} \\
   &  & Spectra Only & Spectra \\
  \hline
  \multicolumn{4}{l}{Photoabsorption} \\
  $N_\mathrm{H}$ & \eVS 10$^{21}\:$cm$^{-2}$ & $3.52\pm0.04$ & $5.4\pm0.4$\footnote{~In the joint fit, photoabsorption was described with \texttt{tbnew} model \mbox{\citep{Wilms2000,JuettWilms2006} and the best-fitting abundances (see \Sect{sec:absorptionEdges}).}} \\
  \hline
  \multicolumn{4}{l}{{(Broken) power-law}} \\
  $\Gamma_1$ (HETGS) & \eVS & $1.51\pm0.01$ & $1.60\pm0.01$ \\
  $\Gamma_1$ (PCA) &  & \nodata & $1.73\pm0.01$ \\
  $E_\mathrm{break}$ & \eVS keV & \nodata & $9.0^{+0.3}_{-1.5}$ \\
  $\Gamma_2$ &  & \nodata & $1.50\pm0.01$ \\
  norm & \eVS s$^{-1}\:$cm$^{-2}\:$keV$^{-1}$ & $1.15\pm0.01$ & $1.33\pm0.03$ \\
  \hline
  \multicolumn{4}{l}{{High-energy cutoff}} \\
  $E_\mathrm{cut}$ & \eVS keV & \nodata & $24^{+2}_{-3}$ \\
  $E_\mathrm{fold}$ & \eVS keV & \nodata & $204\pm9$ \\
  \hline
  \multicolumn{4}{l}{{Disk black body}} \\
  $A_\mathrm{disk}$ (\Equation~\ref{eq:diskbb}) & \eVS $10^3$ & \nodata & $23^{+17}_{-12}$ \\
  $k\,T_\mathrm{col}$ & \eVS keV & \nodata & $0.25^{+0.03}_{-0.02}$ \\
  \hline
  \multicolumn{4}{l}{{Narrow iron K$\alpha$ line}} \\
  $E_{0,\mathrm{narrow}}$ & \eVS keV & \nodata & $6.4001^{+0.0004}_{-0.0092}$ \\
  $\sigma_\mathrm{narrow}$ & \eVS eV (!) & \nodata & $<0.1$ \\
  $A_\mathrm{narrow}$ & \eVS $10^{-3}\:$s$^{-1}\:$cm$^{-2}$\ifApJ{{\color{red}$\:$keV$^{-1}$}}{} & \nodata & $1.0\pm0.3$ \\
  \hline
  \multicolumn{4}{l}{{Broad iron K$\alpha$ line}} \\
  $E_{0,\mathrm{broad}}$ & \eVS keV & \nodata & 6.4 (fixed) \\
  $\sigma_\mathrm{broad}$ & keV & \nodata & $0.6\pm0.1$ \\
  $A_\mathrm{broad}$ & \eVS $10^{-3}\:$s$^{-1}\:$cm$^{-2}$\ifApJ{{\color{red}$\:$keV$^{-1}$}}{} & \nodata & $4.3^{+2.7}_{-0.5}$ \\
  \hline
  \multicolumn{4}{l}{{Relative flux calibration (constant factor)}} \\
  $c_\mathrm{MEG -1}$ & \eVS & 1 (fixed) & 1 (fixed) \\
  $c_\mathrm{MEG +1}$ &  & $1.00\pm0.01$ & $0.99\pm0.01$ \\
  $c_\mathrm{HEG -1}$ &  & $1.04\pm0.01$ & $1.04\pm0.01$ \\
  $c_\mathrm{HEG +1}$ &  & $1.01\pm0.01$ & $1.02\pm0.01$ \\
  $c_\mathrm{PCA}$ &  & \nodata & $1.18\pm0.03$ \\
  $c_\mathrm{HEXTE}$ & \eVS & \nodata & $1.03^{+0.02}_{-0.03}$ \\
  \hline
  \multicolumn{4}{l}{{Pile-up scales}} \\
  $\gamma_\mathrm{MEG -1}$ & \eVS 10$^{-2}\:$s$\;$\AA & $5.6\pm0.1$ & $5.6\pm0.1$ \\
  $\gamma_\mathrm{MEG +1}$ & 10$^{-2}\:$s$\;$\AA & $6.0\pm0.1$ & $5.9\pm0.1$ \\
  $\gamma_\mathrm{HEG -1}$ & 10$^{-2}\:$s$\;$\AA & $4.5\pm0.3$ & $4.6\pm0.3$ \\
  $\gamma_\mathrm{HEG +1}$ & \eVS 10$^{-2}\:$s$\;$\AA & $3.5\pm0.3$ & $3.8\pm0.3$ \\
  \hline
  \multicolumn{4}{l}{{Fit-statistics}} \\
  $\chi^2$ & \eVS & 11745 & 12180 \\
  dof &  & 11274\footnote{~All data have been rebinned to contain $\ge50$ counts\,bin$^{-1}$.}$-$10\footnote{~The model contains actually more parameters, as the absorption lines have already been included (see Tables~\ref{tab:H-He-like}--\ref{tab:H-He-like_columnDensities}).} & $11274^b+293^b-25^c$ \\
  $\chi^2_\mathrm{red}$ & \eVS & 1.04 & 1.06 \\
  \hline
  \multicolumn{4}{l}{{Absorbed flux (pile-up corrected)}} \\
  $S_\mathrm{0.5-10\,keV}$ & \eVS photons\,s$^{-1}$\,cm$^{-2}$ &  \multicolumn{2}{c}{1.4} \\
  $F_\mathrm{0.5-10\,keV}$ & $10^{-9}$\,erg\,s$^{-1}$\,cm$^{-2}$ &  \multicolumn{2}{c}{7.4} \\
  $S_\mathrm{0.5-250\,keV}$ & photons\,s$^{-1}$\,cm$^{-2}$ & \nodata & 1.8 \\
  $F_\mathrm{0.5-250\,keV}$ & \eVS $10^{-9}$ erg\,s$^{-1}$\,cm$^{-2}$ & \nodata & 24 \\
  \hline
  \multicolumn{4}{l}{{Unabsorbed luminosity, assuming $d=2.5\,$kpc \citep{Ninkov1987a}}} \\
  $L_\mathrm{0.5-10\,keV}$ & \eVS $10^{37}$\,erg\,s$^{-1}$ & 0.67 & 0.78 \\
  $L_\mathrm{0.5-250\,keV}$ & \eVS $10^{37}$\,erg\,s$^{-1}$ & \nodata & 3.1 \\
  \hline
 \end{tabular}}\\\ \\
 {\bfseries Notes.}
 Error bars indicate 90\% confidence intervals for one interesting parameter.
\end{table}
The broadband spectrum of \Cyg{} in the hard state
can be described by a broken power-law with exponential cutoff (\Fig{fig:RXTEspectrum}).
Since the parameters of this phenomenological model
are correlated with those of physical Comptonization models
\citep[see, e.g.,][Fig.~11]{Wilms2006},
we are justified in using the aforementioned simple continuum model
for this paper focusing on the spectroscopy of the wind.
We describe the whole \mbox{0.5--250\,keV} spectrum consistently with one broken power-law model,
i.e., there is no need to fit the continuum locally.

Taking pile-up into account (\Sect{sec:pile-up}),
the nondip \Chandra{} spectra alone can already be described quite well
by a weakly absorbed, relatively flat power-law spectrum
with a photon index $\Gamma_1=1.51\pm0.01$ (Table~\ref{tab:continuum})%
\ifApJ{{\color{red}Due to the joint fit with \Chandra, the Fe line consists
       of narrow and a broad component, see \Fig{fig:IronLine}}}{}.
This result is consistent with the fact that the break energy
of the broadband broken power-law spectrum is found at $E_\mathrm{break} = 9\,$keV,
i.e., the \Chandra{} data are virtually entirely in the regime
of the (steeper) photon index $\Gamma_1$.
The onset of the exponential cutoff (with folding energy $E_\mathrm{fold}$)
is at $E_\mathrm{cut}=24\,$keV and thus also well above the spectral range of \Chandra.
As it is known that there are cross-calibration uncertainties
between \Chandra{} and \RXTE{} \citep{Kirsch2005},
we use constant factors $c_i$ for the relative flux calibration of every spectrum
and also separate parameters $\Gamma_1$(HETGS) and $\Gamma_1$(PCA),
for which we find similar values within the joint model (see Table~\ref{tab:continuum}).
In order to describe a weak soft excess, we add a thermal disk component,
which accounts for $\sim$9\,\% of the unabsorbed 0.5--10\,keV flux.
The disk has a color temperature of $kT_\mathrm{col}=0.25\,$keV;
similar to that found by \citet{Makishima2008}
and \citet{BalucinskaChurch1995},
who described the soft excess with a $kT=0.13\pm0.02$\,keV blackbody only,
but the temperature $T_\mathrm{col}$ may be too high by a factor $f\gtrsim1.7$
\citep{ShimuraTakahara1995}.
The norm parameter of the \texttt{diskbb} model is
\begin{equation}
 A_\mathrm{disk} \;\;=\;\; \mathtt{diskbb\ .\ norm} \;\;=\;\; \left(\frac{R_\mathrm{col}/\mathrm{km}}{d/10\,\mathrm{kpc}}\right)^2 \cdot\: \cos\,\theta \;\;,
 \label{eq:diskbb}
\end{equation}
where $d$ is the distance and
$\theta$ is the inclination of the disk,
which can deviate from the orbital inclination $i\approx35^\circ$ \citep{Herrero1995}
as the disk may be precessing with a tilt $\delta\approx37^\circ$ \citep{Brocksopp1999b}.
$R_\mathrm{col}$ is related to the inner radius of the disk, $R_\mathrm{in} = \eta\,g(i)\,f^2 \cdot R_\mathrm{col}$
\citep[with $\eta\approx0.6-0.7$ and $g(i)\approx0.7-0.8$;][]{MerloniFabian2000}.
In spite of these uncertainties and the large statistical error of $A_\mathrm{disk}$ (more than 50\,\%),
the inner disk radius can be estimated to $1.5\,R_\mathrm{S}\lesssim R_\mathrm{in} \lesssim 10\,R_\mathrm{S}$
if a distance $d\approx2.5\,$kpc \citep{Ninkov1987a}
and a Schwarzschild radius $R_\mathrm{S}\approx30\,$km \citep{Herrero1995} are assumed.
Thus, the disk is consistent with extending close to the innermost stable circular orbit (ISCO).
\begin{figure}\centering
 \includegraphics[width=0.95\columnwidth]{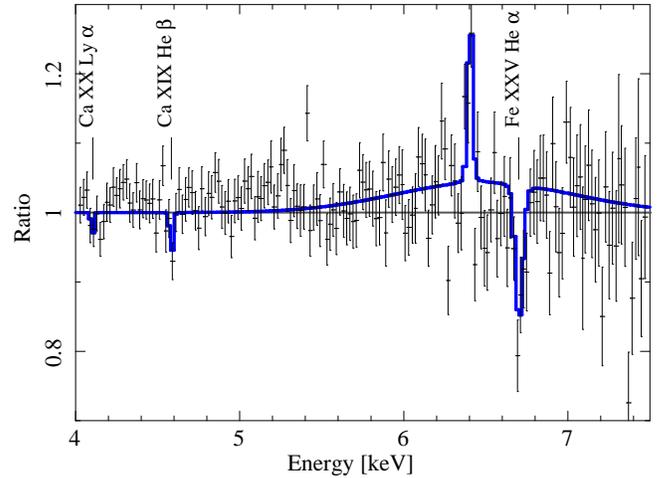}
 \caption{\Chandra-HEG spectrum in the Fe line region.
          The model includes both the narrow and the broad K$\alpha$ emission line,
          the latter as required by the simultaneous \RXTE-PCA data,
          and the \Ion{Fe}{xxv} and \Ion{Ca}{xix}/\textsc{xx} absorption lines.
         }
 \label{fig:IronLine}
\end{figure}

The good S/N afforded by the \RXTE-PCA data
clearly reveals an iron fluorescence K$\alpha$ line.
While the instrumental response of the proportional counters
does not allow for a resolution of the line profile details,
i.e., whether it is narrow or relativistically broadened,
the \Chandra-HEG spectra (\Fig{fig:IronLine})
do resolve a strong narrow component at 6.4\,keV.
Nevertheless, since the integrated flux
of the \Chandra{} measured line is insufficient
to account for all of the PCA residuals in this region,
we include an additional broad feature to our modeling,
which is also compatible with the \Chandra{} spectrum.
Given the relatively low S/N at these energies,
we do not model the broad iron line with a proper physical model
such as a relativistically broadened line,
but use a Gaussian with its energy fixed at 6.4\,keV,
as the latter is hardly constrained by the data.
These results illustrate the synergy of the simultaneous observation
with complementary instruments,
as the combination of narrow and broad line could only be
revealed by the analysis with a joint model.

Our global model for the continuum spectrum now enables us
to address the features of the high-resolution \Chandra{} spectra,
which is the topic in the remainder of this section.

\subsection{Neutral Absorption} \label{sec:absorptionEdges}
\begin{figure}\centering
 \includegraphics[width=0.95\columnwidth]{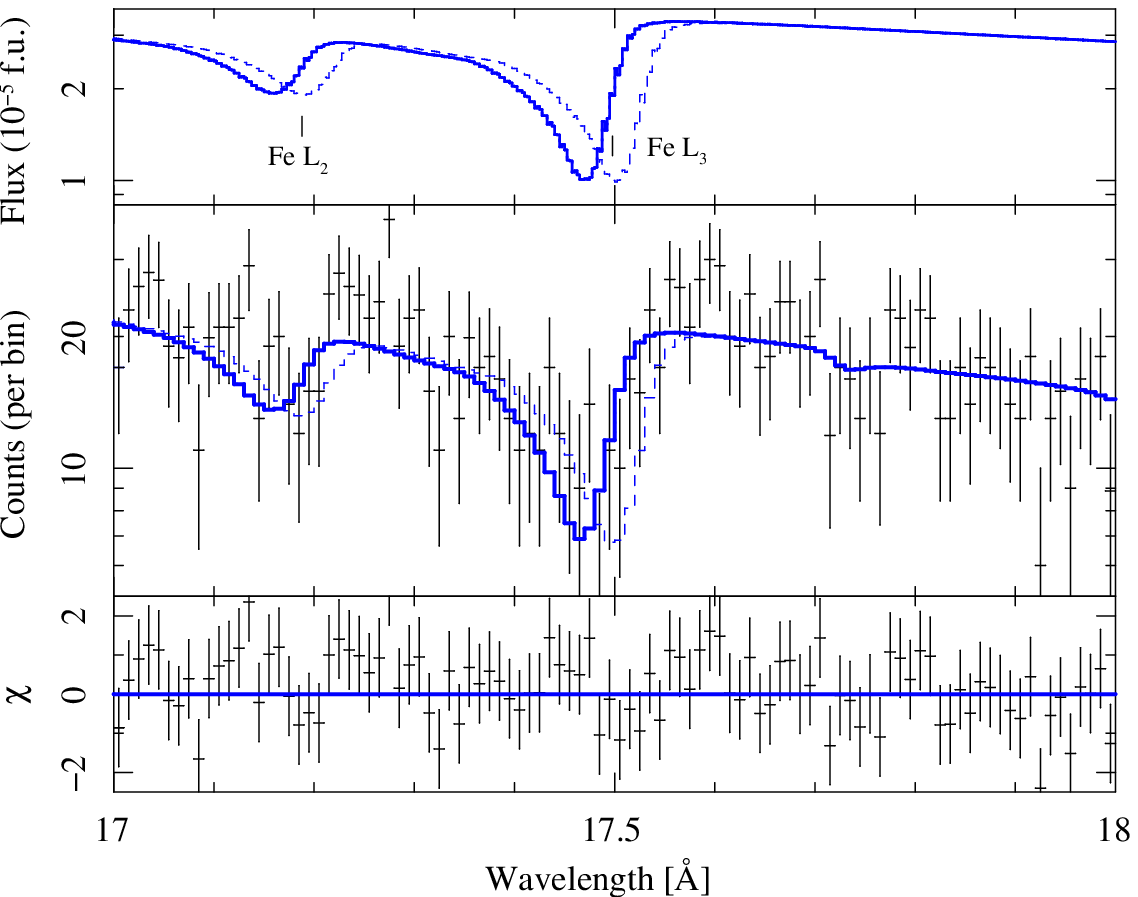}
 \caption{Spectral region around the Fe L-edges.
          The top panel shows the model flux in units of $10^{-5}$\,ph.\,s$^{-1}$\,cm$^{-2}$\,\AA$^{-1}$ (solid line).
          The dashed lines describe the unshifted model, see the text.
          The second panel shows the count rate of all the four HETGS spectra (MEG$\pm1$, HEG$\pm1$),
          which have been combined with a resolution of 10\,m\AA{}, and the folded model.
          The bottom panel displays the residuals $\chi=($data--model)/error.
         }
 \label{fig:FeEdge}
 \includegraphics[width=0.95\columnwidth]{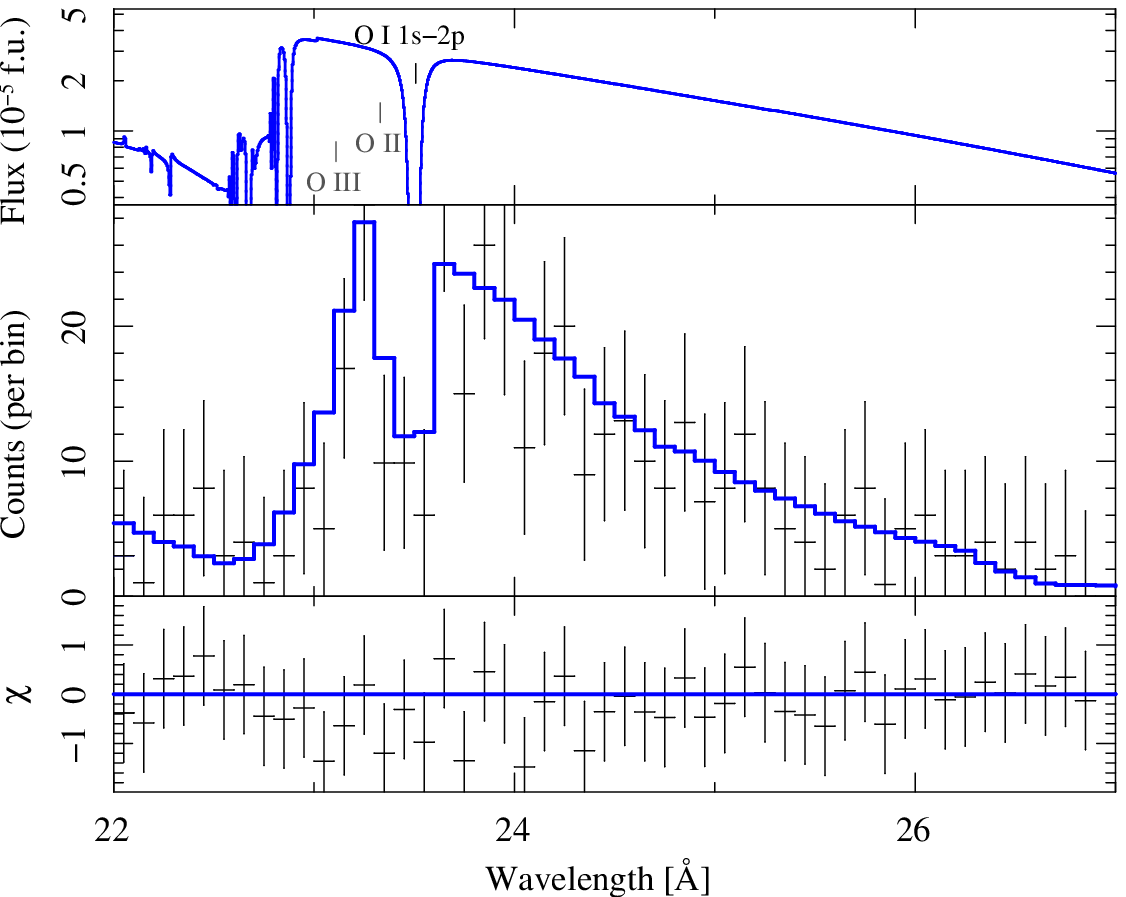}
 \caption{Same as in \Fig{fig:FeEdge}, but for the spectral region around the O K-edge.
          Both MEG$\pm1$ spectra have been combined with a resolution of 0.1\,\AA{}.
         }
 \label{fig:Oedge}
 \includegraphics[width=0.95\columnwidth]{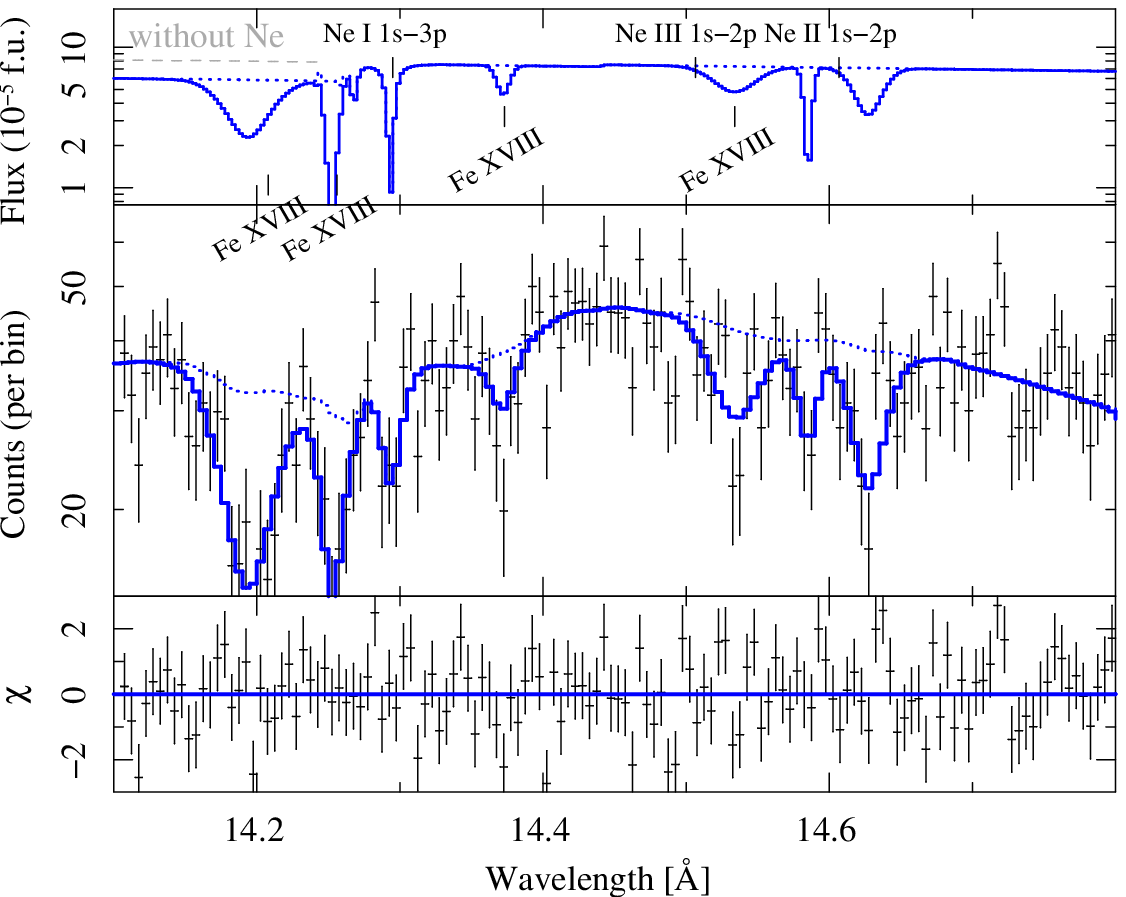}
 \caption{Same as in \Fig{fig:FeEdge}, but for the spectral region around the Ne K-edge.
          All spectra have been combined with a resolution of 5\,m\AA{}.
          The dotted lines show the absorbed continuum model without additional absorption lines.
          The light dashed line shows the model without the absorption of Ne.
         }
 \label{fig:NeEdge}
\end{figure}
Absorbing columns can be measured most accurately
from the discrete edges in high-resolution spectra
at the ionization thresholds.
We detect the most prominent L-shell absorption edge of iron 
and the K-shell absorption edges of oxygen 
and neon. 
\citet{Juett2004,Juett2006} have inferred the fine structure
at those edges from earlier \Chandra-HETGS observations of \Cyg{}
and other bright X-ray binaries:
the Fe L$_2$ and L$_3$ edges,
due to the ionization of a 2p$_{1/2}$ or a 2p$_{3/2}$ electron, respectively,
are separately detectable at 17.2\,\AA{} and 17.5\,\AA{}.
The O K-edge at 22.8\,\AA{} is accompanied by (1s$\rightarrow$2p) K$\alpha$
and higher ($1$s$\rightarrow n$p) resonance absorption lines.
The K$\alpha$ line occurs at 23.5\,\AA{} for neutral \Ion{O}{i}
and at lower wavelengths for ionized oxygen.
In the case of neon, neutral atoms have closed L-shells,
such that \Ion{Ne}{i} only shows a (1s$\rightarrow$3p) K$\beta$ absorption line
close to the K-edge at 14.3\,\AA{},
while ionized neon also shows K$\alpha$ absorption lines,
e.g., \Ion{Ne}{ii} at 14.6\,\AA{} and \Ion{Ne}{iii} at 14.5\,\AA{}.
Improved modeling of the neutral absorption that takes these features into account
has recently been included in the photoabsorption model
\texttt{tbnew}\footnote{The code for \texttt{tbnew} is available online
 at \href{http://pulsar.sternwarte.uni-erlangen.de/wilms/research/tbabs/}{http://pulsar.sternwarte.uni-} \href{http://pulsar.sternwarte.uni-erlangen.de/wilms/research/tbabs/}{erlangen.de/wilms/research/tbabs/}
 for beta testing and will soon be released.}
\citep{JuettWilms2006},
an extension of the commonly used \texttt{tbvarabs} model \citep{Wilms2000}.
\begin{table*}\centering
 \begin{minipage}{\textwidth}
 \caption{Column Density $N$ and Abundance $A=N/N_\mathrm{H}$ of Neutral Absorbers Detected Along the Line of Sight Toward \Cyg}
 \label{tab:neutralAbundances}
 \begin{tabular}{cc@{\quad\quad}ccc@{\quad\quad}cccc}
  \hline
  \hline
   &  &  &  &  & \multicolumn{4}{c}{Analysis by} \\
   \cline{6-9}
   &  & \multicolumn{3}{c}{\eVS This work} & \citet{Schulz2002} & \multicolumn{3}{c}{\citet{Juett2004,Juett2006}} \\
   &  & \multicolumn{3}{c}{} & ---{}---{}---{}---{}---{}---{}--- & \multicolumn{3}{c}{---{}---{}---{}---{}---{}---{}---{}---{}---{}---{}---{}---{}---{}---{}---{}---{}---{}---{}---{}---{}---{}---{}} \\
   &  & \multicolumn{3}{c}{ObsID 3814} & ObsID 107 & ObsID 107 & ObsID 3407 & ObsID 3724 \\
   &  & \multicolumn{3}{c}{($L_{0.5-10\,\mathrm{kev}}=0.7\!\times\!10^{37}\,$erg\,s$^{-1}$)} & \multicolumn{2}{c}{($L_{0.5-10\,\mathrm{kev}}=1.6\!\times\!10^{37}\,$erg\,s$^{-1}$)} & ($3.1\!\times\!10^{37}\,$erg\,s$^{-1}$) & (soft state) \\
   &  & \multicolumn{3}{c}{---{}---{}---{}---{}---{}---{}---{}---{}---{}---{}---{}---{}---{}---{}---{}---{}---{}---{}---{}---{}} & ---{}---{}---{}---{}---{}---{}--- & ---{}---{}---{}---{}--- & ---{}---{}---{}---{}--- & ---{}---{}---{}---{}--- \\
  Element & $12 + \log\,A_\mathrm{element}^\mathrm{ISM}$\footnote{
 We use elemental abundances in the interstellar medium, $A_\mathrm{element}^\mathrm{ISM}$ according to \citet{Wilms2000},
 i.e., \texttt{xspec\_abund("wilm");} in \textsc{ISIS}.
} & \eVS $N_\mathrm{element} / N_\mathrm{Fe}$ & $A_\mathrm{element}/A_\mathrm{element}^\mathrm{ISM}$\footnote{
 The relative elemental abundance, $A_\mathrm{element}/A_\mathrm{element}^\mathrm{ISM}$, 
 is a parameter of the \texttt{tbnew} model.
} & $N_\mathrm{element}$\footnote{
 $N_\mathrm{element}$ is the product of $A_\mathrm{element}$ and $N_\mathrm{H}$ (see Table~\ref{tab:continuum}).
} & $N_\mathrm{element}$\footnote{
 \citet{Schulz2002} use the cross sections from \citet{Verner1993}, and \citet{KortrightKim2000} for Fe.
} & $N_\mathrm{element}$\footnote{
 \citet{Juett2004,Juett2006} use the cross section of \citet{GorczycaMcLaughlin2000} for O,
 \citet{Gorczyca2000} for Ne, and \citet{KortrightKim2000} for Fe.
} & ${N_\mathrm{element}}^e$ & ${N_\mathrm{element}}^e$ \\
   &  &  &  & \eVS ($10^{17}\,$cm$^{-2}$) & ($10^{17}\,$cm$^{-2}$) & ($10^{17}\,$cm$^{-2}$) & ($10^{17}\,$cm$^{-2}$) & ($10^{17}\,$cm$^{-2}$) \\
  \hline
  O  \eVS & 8.69 & $15.4\pm0.3$ & $1.45\pm0.01$ & $38.1^{+0.4}_{-0.3}$ & $39.2\pm2.3$ & $63\pm14$ & $24\pm3$ & $29\pm3$ \\
  Ne \eVS & 7.94 & $3.91\pm0.18$ & $2.07^{+0.04}_{-0.03}$ & $9.6\pm0.2$ & $9.43\pm0.32$ & $7.1\pm1.0$ & $7.4^{+0.7}_{-0.3}$ & $8.6\pm0.7$ \\
  Na \eVS & 6.16 & $1.26\pm0.15$ & $40\pm2$ & $3.1\pm0.2$ & \nodata & \nodata & \nodata & \nodata \\
  Mg \eVS & 7.40 & $2.16\pm0.13$ & $4.0\pm0.1$ & $5.3\pm0.2$ & $3.7\pm1.3$ & \nodata & \nodata & \nodata \\
  Al \eVS & 6.33 & $<0.13$ & $<1.8$ & $<0.2$ & \nodata & \nodata & \nodata & \nodata \\
  Si \eVS & 7.27 & $1.45\pm0.17$ & $3.8\pm0.3$ & $3.8\pm0.3$ & $2.3\pm1.8$ & \nodata & \nodata & \nodata \\
   S \eVS & 7.09 & $0.6\pm0.2$ & $2.4^{+0.5}_{-0.3}$ & $1.6^{+0.3}_{-0.2}$ & \nodata & \nodata & \nodata & \nodata \\
  Ar \eVS & 6.41 & $<0.06$ & $<1.0$ & $<0.1$ & \nodata & \nodata & \nodata & \nodata \\
  Ca \eVS & 6.20 & $<0.05$ & $<1.3$ & $<0.1$ & \nodata & \nodata & \nodata & \nodata \\
  Cr \eVS & 5.51 & $<0.18$ & $1.3^{+2.7}_{-1.3}$ & $0.02^{+0.05}_{-0.02}$ & \nodata & \nodata & \nodata & \nodata \\
  Fe \eVS & 7.43 & 1 & $1.75\pm0.03$ & $2.52\pm0.04$ & $1.5\pm0.1$ & $1.47^{+0.25}_{-0.19}$ & $1.06^{+0.07}_{-0.11}$ & $0.96\pm0.09$ \\
  \hline
 \end{tabular}\\~\\
 {\bfseries Notes.}
 \end{minipage}
\end{table*}

As part of the spectral model for the whole continuum,
the \texttt{tbnew} model can be used to describe the absorption edges
detected with the \Chandra{} observation of \Cyg{} discussed in this paper.
Figure~\ref{fig:FeEdge} shows the Fe L-edges
requiring a blueshift by $\Delta\lambda/\lambda_0\cdot c = (540\pm230)\:$km\,s$^{-1}$
of the \texttt{tbnew} model, which relies on
the cross sections of metallic iron measured by \citet{KortrightKim2000}.
Unlike \citet{Schulz2002}, \citet{Miller2005} and 
\citet{Juett2006} have also found that the Fe L$_3$ edge requires a small shift;
their mean position of maximum optical depth is 
$\lambda_\mathrm{Fe\:L_3}=17.498\,$\AA,
but our value $\lambda_\mathrm{Fe\:L_3}=(17.469\pm0.014)\,$\AA{} is still lower.
The shift could be caused by the Doppler effect due to a moving absorber,
by a modified ionization threshold due to chemical bonds,
or ionization of the iron atoms \citep{vanAkenLiebscher2002}.
In an analysis of the \Cyg{} high/soft and low/hard state,
focused on this spectral region,
J.~Lee et al.~(2008, in preparation) find that the Fe L-edges here can likely be modeled
by a heterogeneous combination of gas and condensed matter of iron
in combination with oxygen local to the source environment.
If, as suggested by these authors,
the magnitude of the shift is due to molecules and/or dust,
this shift is one identifying signature of the composition 
and charge state of the condensed state material.
Such direct \Chandra{} X-ray spectroscopic detection of dust
via its associated edge structure
was first suggested for observations of the Fe L-edge
in the active galactic nucleus \mbox{MCG$-$6-30-15} \citep{Lee2001}
and for the observations of the Si K-edge
in the microquasar GRS~1915+105 \citep{Lee2002}.
The latter study associated the observed Si K-edge structure with SiO$_2$,
although the origin\ifApJ{---}{ -- }source environment or \Chandra{} CCD gate structure\ifApJ{---}{ --}
was unclear in this case.

The blueshift of \mbox{$(100\pm130)\:$km\,s$^{-1}$}
which has been determined for all other absorption edges 
is consistent with zero.
The O K-edge can only be seen in the MEG spectra
after heavy rebinning (\Fig{fig:Oedge}).
Nevertheless, the K$\alpha$ resonance absorption line of \Ion{O}{i} is clearly detected.
The region around the Ne edge (\Fig{fig:NeEdge}) is dominated
by \Ion{Fe}{xviii} absorption lines,
possibly blending with the K$\alpha$ absorption lines of \Ion{Ne}{ii} and \Ion{Ne}{iii}.
No other strong edges are clearly visible in the spectrum.
The Na K-edge \citep[at 11.5\,\AA;][]{VernerYakovlev1995}
blends with an absorption line
due to the 2s$\rightarrow$3p excitation of \Ion{Fe}{xxii}.
The Mg K-edge (at 9.5\,\AA) blends
with the \Ion{Ne}{x} Ly\,$\delta$ absorption line.
The Si K-edge (at 6.7\,\AA) is strongly affected by pile-up
and blends with the \Ion{Mg}{xii} Ly\,$\beta$ absorption line.
The S K-edge (at 5.0\,\AA) is relatively weak.
Neutral absorption from these elements is nevertheless required
within the \verb|tbnew| model.

The results for the individual abundances, $A_i=N_i/N_\mathrm{H}$,
and resulting column densities, $N_i$, are presented in Table~\ref{tab:neutralAbundances}.
The alpha process elements O, Ne, Mg, Si, and S are overabundant
with respect to the interstellar medium (ISM) abundances as summarized by \citet{Wilms2000}
and therefore suggest an origin in the system itself.
The total column densities are also compared
with the values obtained by \citet{Schulz2002} and \citet{Juett2004,Juett2006}
from other \Chandra{} observations of \Cyg{}.
Table~\ref{tab:neutralAbundances} includes the corresponding source luminosities
if they are reported in the literature (see also \Sect{sec:prevObs}).
The X-ray flux was highest during the soft state\ifApJ{}{ observation with the}
observation identification (ObsID) 3724.
The column densities confirm the conjecture of \citet{Juett2004}
that a higher (soft) X-ray flux ionizes material local to the \Cyg{} system
and reduces the neutral abundances.

The inferred hydrogen column density $N_\mathrm{H}$ (see Table~\ref{tab:continuum})
is in very good agreement with that from \textsl{ASCA} observations during the soft state in 1996,
namely $N_\mathrm{H}=(5.3\pm0.2)\times10^{21}\,$cm$^{-2}$ \citep{Dotani1997},
and also with $N_\mathrm{H}=6.2\!\times\!10^{21}\:\mathrm{cm}^{-2}$
obtained from two other different \Chandra{} observations
\citep[see also \Sect{sec:prevObs}]{Schulz2002,Miller2002}.
We note that the large column density toward \Cyg{} 
found by many online tools, $(7.2-7.8)\times\!10^{21}\,$cm$^{-2}$,
is obtained from a coarse grid\ifApJ{---}{ -- }with (0.675\,deg)$^2$ pixel size%
\ifApJ{---}{ -- }of $N_\mathrm{H}$ measurements at 21\,cm \citep{Kalberla2005},
which does not resolve the strong variations of $N_\mathrm{H}$
in the region around \Cyg{} \citep{Russell2007}.

\begin{figure*}\centering
 \includegraphics[width=0.95\textwidth]{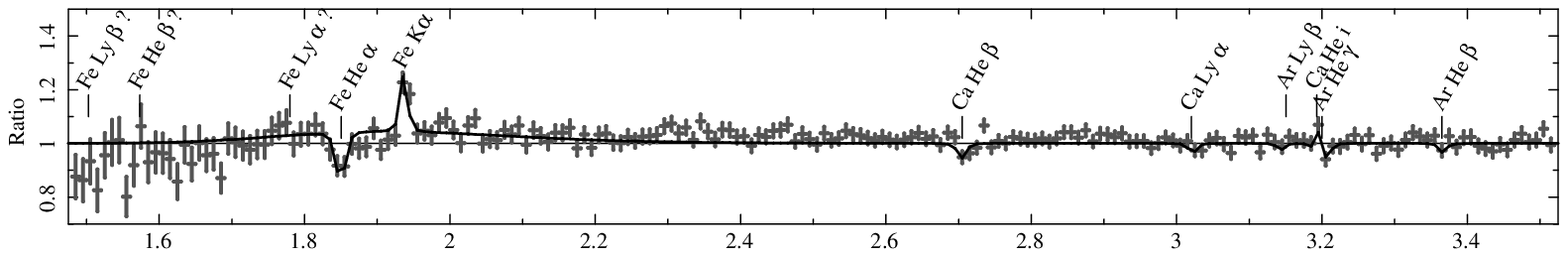}\\[0.01\textwidth]
 \includegraphics[width=0.95\textwidth]{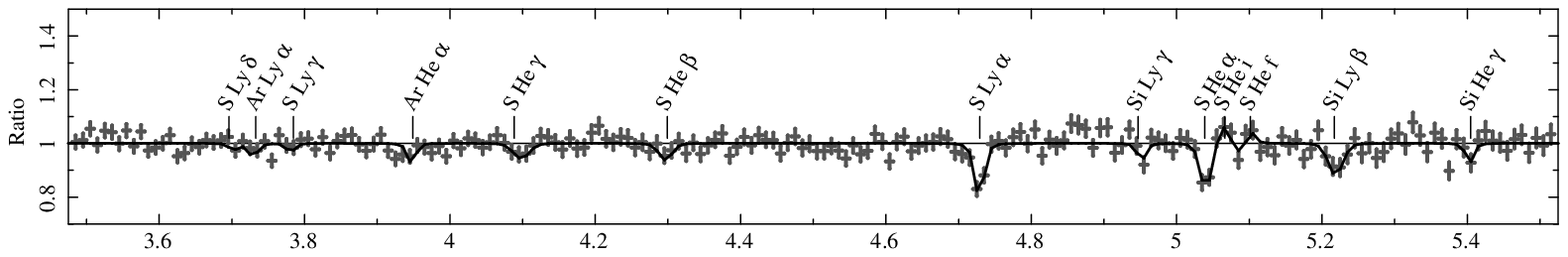}\\[0.01\textwidth]
 \includegraphics[width=0.95\textwidth]{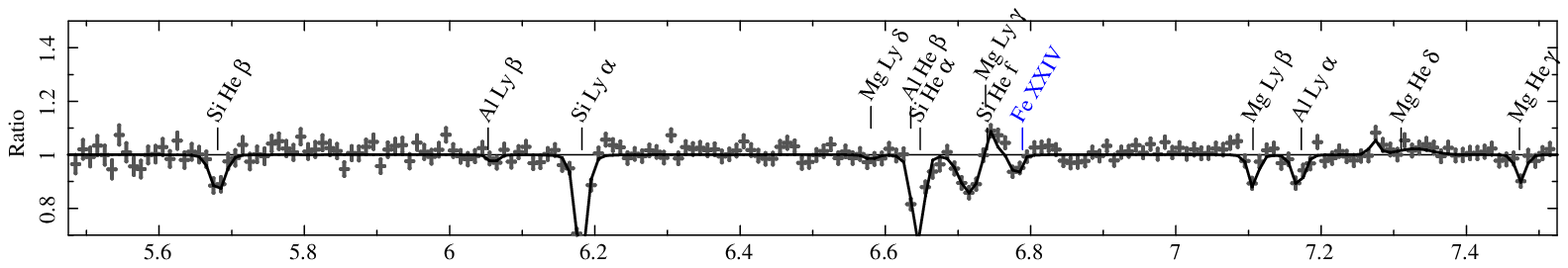}\\[0.01\textwidth]
 \includegraphics[width=0.95\textwidth]{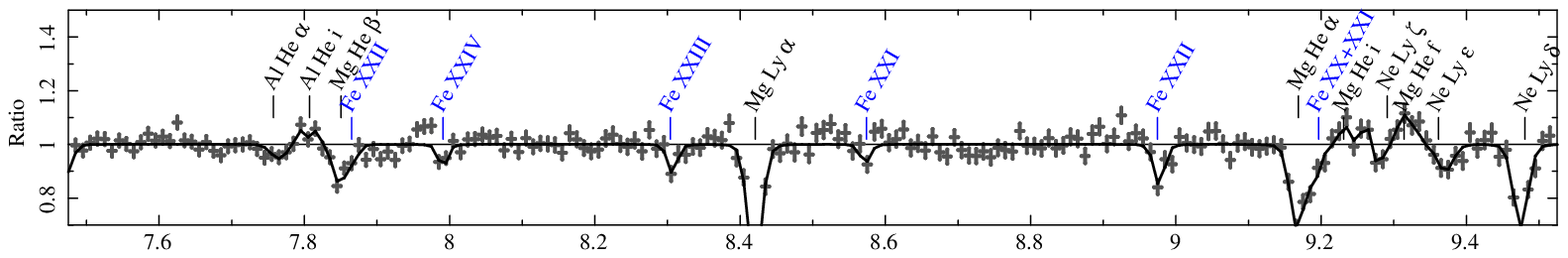}\\[0.01\textwidth]
 \includegraphics[width=0.95\textwidth]{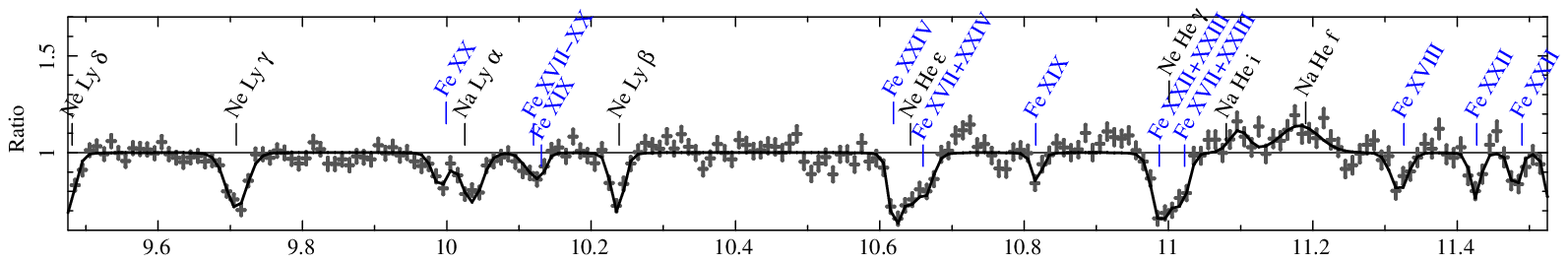}\\[0.01\textwidth]
 \includegraphics[width=0.95\textwidth]{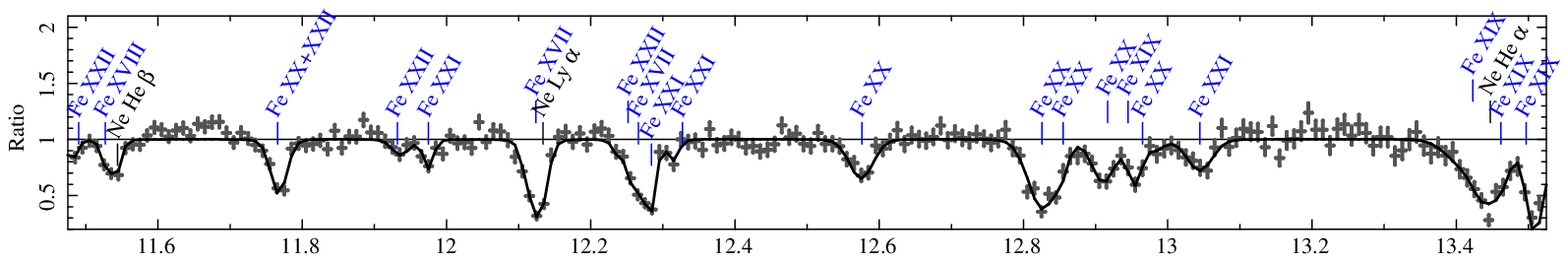}\\[0.01\textwidth]
 \includegraphics[width=0.95\textwidth]{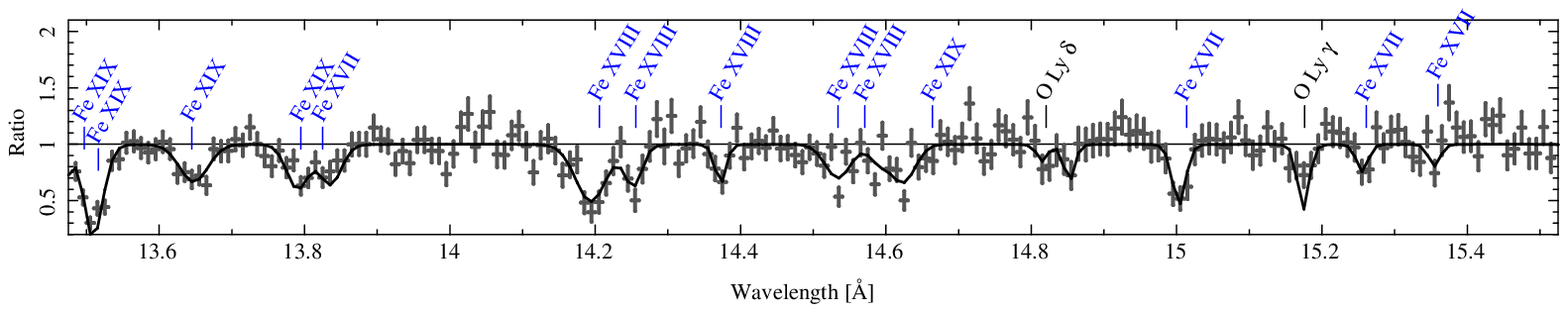}
 \caption{\Chandra{} (nondip) spectrum of \Cyg{},
          shown as ratio of data and continuum model (Table~\ref{tab:continuum}).
          For visual clarity, the data have been rebinned
          to a common resolution of 10\,m\AA{},
          and all MEG$\pm1$ and HEG$\pm1$ spectra have been combined.
          The Gaussian line fits and identifications are shown as well.
          (The labels mark the lines' rest wavelengths.)
          Iron L-shell transitions (lines of \Ion{Fe}{$\small <$xxv})
          are shown in blue in the online journal.
         }
 \label{fig:spectrum}
\end{figure*}
\subsection{Absorption Lines of H- and He-Like Ions} \label{sec:absorptionLines}
\newcommand{\defTwoLine}[1]{\underline{\textbf{#1}}}
\newcommand{\defDetLine}[1]{\textbf{#1}}
\newcommand{\unsureLine}[1]{#1}
\newcommand{\notDetLine}[1]{(#1)}
\newcommand{\notCovLine}[1]{[#1]}
\begin{table*}{\centering
 \caption{Overview on the Detected Lines from H- and He-Like Ions: Theoretical Rest Wavelengths in \AA{}}
 \label{tab:H-He-like}
 \begin{tabular}{c@{$\;\;$}l@{\quad}ccccccccccc}
  \hline
  \hline
   \multicolumn{2}{c}{Transition}                                                     & O                  & Ne                 & Na                 & Mg                & Al                & Si                & S                 & Ar                & Ca                & Fe                & Ni            \\
   \multicolumn{2}{r}{Hydrogen-like (1 electron)}                                     &   \sc viii         &    \sc x           &    \sc xi          &    \sc xii        &    \sc xiii       &    \sc xiv        &   \sc xvi         &    \sc xviii      &    \sc xx         &    \sc xxvi       &    \sc xxviii \\
  \hline
   Ly\,$\alpha$   & 1s$\:(^2\mathrm{S}_{1/2})\rightarrow2$p$\:(^2\mathrm{P}_{3/2,1/2})$ & \defDetLine{18.97} & \defDetLine{12.13} & \defDetLine{10.03} & \defDetLine{8.42} & \defDetLine{7.17} & \defDetLine{6.18} & \defDetLine{4.73} & \defDetLine{3.73} & \defDetLine{3.02} & \notDetLine{1.78} & \defDetLine{1.53} \\
   Ly\,$\beta$    & 1s$\:(^2\mathrm{S}_{1/2})\rightarrow3$p$\:(^2\mathrm{P}_{3/2,1/2})$ & \defDetLine{16.01} & \defDetLine{10.24} & \unsureLine{ 8.46} & \defDetLine{7.11} & \unsureLine{6.05} & \defDetLine{5.22} & \unsureLine{3.99} & \defDetLine{3.15} & \notDetLine{2.55} & \unsureLine{1.50} & \unsureLine{1.29} \\
   Ly\,$\gamma$   & 1s$\:(^2\mathrm{S}_{1/2})\rightarrow4$p$\:(^2\mathrm{P}_{3/2,1/2})$ & \defDetLine{15.18} & \defDetLine{ 9.71} & \unsureLine{ 8.02} & \notDetLine{6.74} & \unsureLine{5.74} & \defDetLine{4.95} & \defDetLine{3.78} & \unsureLine{2.99} & \unsureLine{2.42} & \unsureLine{1.43} & \defDetLine{1.23} \\
   Ly\,$\delta$   & 1s$\:(^2\mathrm{S}_{1/2})\rightarrow5$p$\:(^2\mathrm{P}_{3/2,1/2})$ & \defDetLine{14.82} & \defDetLine{ 9.48} & \unsureLine{ 7.83} & \unsureLine{6.58} & \notDetLine{5.60} & \notDetLine{4.83} & \defDetLine{3.70} & \notDetLine{2.92} & \notDetLine{2.36} & \notDetLine{1.39} & \notDetLine{1.20} \\
   Ly\,$\epsilon$ & 1s$\:(^2\mathrm{S}_{1/2})\rightarrow6$p$\:(^2\mathrm{P}_{3/2,1/2})$ & \notDetLine{14.63} & \defDetLine{ 9.36} & \notDetLine{ 7.73} & \unsureLine{6.50} & \notDetLine{5.53} & \notDetLine{4.77} & \notDetLine{3.65} & \notDetLine{2.88} & \notDetLine{2.33} & \notDetLine{1.37} \\
   Ly\,$\zeta$    & 1s$\:(^2\mathrm{S}_{1/2})\rightarrow7$p$\:(^2\mathrm{P}_{3/2,1/2})$ & \notDetLine{14.52} & \unsureLine{ 9.29} & \notDetLine{ 7.68} & \unsureLine{6.45} & \notDetLine{5.49} & \notDetLine{4.73} & \notDetLine{3.62} & \notDetLine{2.86} & \notDetLine{2.31} & \notDetLine{1.36} \\
   Ly\,$\eta$     & 1s$\:(^2\mathrm{S}_{1/2})\rightarrow8$p$\:(^2\mathrm{P}_{3/2,1/2})$ & \notDetLine{14.45} & \unsureLine{ 9.25} & \notDetLine{ 7.64} & \notDetLine{6.42} & \notDetLine{5.47} & \notDetLine{4.71} & \notDetLine{3.60} & \notDetLine{2.85} & \notDetLine{2.30} & \notDetLine{1.36} \\
   Ly\,$\theta$   & 1s$\:(^2\mathrm{S}_{1/2})\rightarrow9$p$\:(^2\mathrm{P}_{3/2,1/2})$ & \notDetLine{14.41} & \notDetLine{ 9.22} & \notDetLine{ 7.61} & \notDetLine{6.40} & \notDetLine{5.45} & \notDetLine{4.70} & \notDetLine{3.59} & \notDetLine{2.84} & \notDetLine{2.29} & \notDetLine{1.35} \\
   $\cdots$ \\
   limit          & $1$s$\:(^2\mathrm{S}_{1/2})\rightarrow\infty$                       & \notDetLine{14.23} & \notDetLine{ 9.10} & \notDetLine{ 7.52} & \unsureLine{6.32} & \unsureLine{5.38} & \notDetLine{4.64} & \notDetLine{3.55} & \notDetLine{2.80} & \notDetLine{2.27} & \unsureLine{1.34} & \notDetLine{1.15} \\
  \hline
  \ \\
  \ \\
  \hline
  \hline
   \multicolumn{2}{c}{Transition}                                                      & O                  & Ne                 & Na                 & Mg                & Al                & Si                & S                 & Ar                & Ca                & Fe                & Ni                \\
   \multicolumn{2}{r}{Helium-like (2 electrons)}                                       &   \sc vii          &    \sc ix          &    \sc x           &    \sc xi         &    \sc xii        &    \sc xiii       &   \sc xv          &    \sc xvii       &    \sc xix        &    \sc xxv        &    \sc xxvii      \\
  \hline
   f [em.]                 & $1$s$^2\:(^1\mathrm{S}_0)\leftarrow 1$s$2$s$\:(^3\mathrm{S}_1)$ & \unsureLine{22.10} & \notDetLine{13.70} & \unsureLine{11.19} & \defTwoLine{9.31} & \unsureLine{7.87} & \defDetLine{6.74} & \defDetLine{5.10} & \notDetLine{3.99} & \notDetLine{3.21} & \notDetLine{1.87} \\
   i [em.]              & $1$s$^2\:(^1\mathrm{S}_0)\leftarrow1$s$2$p$\:(^3\mathrm{P}_{1,2})$ & \unsureLine{21.80} & \notDetLine{13.55} & \unsureLine{11.08} & \defTwoLine{9.23} & \defTwoLine{7.81} & \notDetLine{6.69} & \defDetLine{5.07} & \unsureLine{3.97} & \defDetLine{3.19} & \notDetLine{1.86} & \unsureLine{1.60} \\
   \hline
   r $\equiv$ He\,$\alpha$ & $1$s$^2\:(^1\mathrm{S}_0)\rightarrow1$s$2$p$\:(^1\mathrm{P}_1)$ & \unsureLine{21.60} & \defDetLine{13.45} & \defDetLine{11.00} & \defDetLine{9.17} & \defDetLine{7.76} & \defDetLine{6.65} & \defDetLine{5.04} & \defDetLine{3.95} & \defDetLine{3.18} & \defDetLine{1.85} & \unsureLine{1.59} \\
   He\,$\beta$             & $1$s$^2\:(^1\mathrm{S}_0)\rightarrow1$s$3$p$\:(^1\mathrm{P}_1)$ & \unsureLine{18.63} & \defDetLine{11.54} & \unsureLine{ 9.43} & \defDetLine{7.85} & \defDetLine{6.64} & \defDetLine{5.68} & \defDetLine{4.30} & \defDetLine{3.37} & \defDetLine{2.71} & \unsureLine{1.57} & \notDetLine{1.35} \\
   He\,$\gamma$            & $1$s$^2\:(^1\mathrm{S}_0)\rightarrow1$s$4$p$\:(^1\mathrm{P}_1)$ & \notDetLine{17.77} & \defDetLine{11.00} & \unsureLine{ 8.98} & \defDetLine{7.47} & \unsureLine{6.31} & \defDetLine{5.40} & \defDetLine{4.09} & \defDetLine{3.20} & \notDetLine{2.57} & \unsureLine{1.50} & \notDetLine{1.28} \\
   He\,$\delta$            & $1$s$^2\:(^1\mathrm{S}_0)\rightarrow1$s$5$p$\:(^1\mathrm{P}_1)$ & \notDetLine{17.40} & \unsureLine{10.77} & \unsureLine{ 8.79} & \unsureLine{7.31} & \notDetLine{6.18} & \notDetLine{5.29} & \unsureLine{4.00} & \notDetLine{3.13} & \notDetLine{2.51} & \unsureLine{1.46} & \notDetLine{1.25} \\
   He\,$\epsilon$          & $1$s$^2\:(^1\mathrm{S}_0)\rightarrow1$s$6$p$\:(^1\mathrm{P}_1)$ & \notDetLine{17.20} & \defDetLine{10.64} & \notDetLine{ 8.69} & \notDetLine{7.22} & \notDetLine{6.10} & \unsureLine{5.22} & \unsureLine{3.95} & \notDetLine{3.10} \\
   He\,$\zeta$             & $1$s$^2\:(^1\mathrm{S}_0)\rightarrow1$s$7$p$\:(^1\mathrm{P}_1)$ & \notDetLine{17.09} & \unsureLine{10.56} & \notDetLine{ 8.63} & \notDetLine{7.17} & \notDetLine{6.06} & \notDetLine{5.19} & \notDetLine{3.92} \\
   He\,$\eta$              & $1$s$^2\:(^1\mathrm{S}_0)\rightarrow1$s$8$p$\:(^1\mathrm{P}_1)$ & \notDetLine{17.01} & \notDetLine{10.51} & \notDetLine{ 8.59} & \notDetLine{7.14} & \notDetLine{6.03} & \notDetLine{5.16} & \notDetLine{3.90} \\
   $\cdots$ \\
   limit                   & $1$s$^2\:(^1\mathrm{S}_0)\rightarrow1$s$\:\infty$             & \unsureLine{16.77} & \unsureLine{10.37} & \unsureLine{ 8.46} & \notDetLine{7.04} & \unsureLine{5.94} & \unsureLine{5.09} & \notDetLine{3.85} & \notDetLine{3.01} & \unsureLine{2.42} & \unsureLine{1.40} & \notDetLine{1.20} \\
  \hline
 \end{tabular}\\
 }~\\
 {\bfseries Notes.}
 Lines with (wavelengths in parentheses) are not detected in our \Chandra-HETGS observation of \Cyg,
 while lines indicated with \textbf{bold wavelengths} are clearly detected
 and those with \defTwoLine{underlined wavelengths} are detected as two components.
 The wavelengths of the lines are taken from the CXC atomic database \textsc{atomdb} and the table of \citet{Verner1996},
 those of the series limits (=~K-ionization thresholds) are from \citet{VernerYakovlev1995}.
\end{table*}
\begin{table*}{\centering
 \caption{Results from the Detected Absorption Lines from H- and He-Like Ions: Velocity Shifts $v=(\lambda-\lambda_0)/\lambda_0\cdot c$ in $\rm km\,s^{-1}$}
 \label{tab:H-He-like_velocityShifts}
 \begin{tabular}{c@{\quad}cccccccccccc}
  \hline
  \hline
          & O                  & Ne                 & Na                 & Mg                & Al                & Si                & S                 & Ar                & Ca                & Fe          \\
  H-like  &   \sc viii         &    \sc x           &    \sc xi          &    \sc xii        &    \sc xiii       &    \sc xiv        &   \sc xvi         &    \sc xviii      &    \sc xx         &    \sc xxvi \\
  \hline
  \eVS Ly $\alpha$ & $-718^{+1716}_{-281}$ & $-128^{+35}_{-27}$ & $284^{+81}_{-81}$ & $-34^{+26}_{-34}$ & $-171^{+88}_{-104}$ & $-60^{+32}_{-33}$ & $-43^{+114}_{-95}$ & $-322^{+712}_{-617}$ & $164^{+838}_{-402}$ & \nodata\\
  \eVS Ly $\beta$ & $-90^{+187}_{-185}$ & $-72^{+50}_{-47}$ & \nodata & $17^{+27}_{-73}$ & $344^{+431}_{-448}$ & $75^{+263}_{-238}$ & \nodata & $-750^{+710}_{-451}$ & \nodata & \nodata\\
  \eVS Ly $\gamma$ & $-39^{+85}_{-81}$ & $7^{+64}_{-64}$ & \nodata & \nodata & \nodata & $308^{+333}_{-193}$ & $-347^{+393}_{-399}$ & \nodata & \nodata & \nodata\\
  \eVS Ly $\delta$ & $-109^{+1109}_{-198}$ & $-192^{+46}_{-46}$ & \nodata & $3^{+922}_{-1401}$ & \nodata & \nodata & $420^{+536}_{-489}$ & \nodata & \nodata & \nodata\\
  \eVS Ly $\epsilon$ & \nodata & $288^{+213}_{-177}$ & \nodata & \nodata & \nodata & \nodata & \nodata & \nodata & \nodata & \nodata\\
  \hline
  \eVS Ly series & $-41^{+67}_{-72}$ & $-97^{+28}_{-20}$ & $207^{+104}_{-101}$ & $-38^{+27}_{-28}$ & $-183^{+140}_{-117}$ & $-54^{+33}_{-32}$ & $-90^{+74}_{-73}$ & $-327^{+41}_{-128}$ & \nodata & \nodata \\
  \hline
  \ \\
  \ \\
  \hline
  \hline
          & O                  & Ne                 & Na                 & Mg                & Al                & Si                & S                 & Ar                & Ca                & Fe          \\
  He-like &   \sc vii          &    \sc ix          &    \sc x           &    \sc xi         &    \sc xii        &    \sc xiii       &   \sc xv          &    \sc xvii       &    \sc xix        &    \sc xxv  \\
  \hline
  \eVS He $\alpha$ & \nodata & $-203^{+85}_{-91}$ & \nodata & $-71^{+35}_{-49}$ & $286^{+187}_{-574}$ & $-30^{+23}_{-46}$ & $84^{+127}_{-138}$ & $-142^{+250}_{-233}$ & $1124^{+98}_{-1069}$ & $-116^{+501}_{-531}$\\
  \eVS He $\beta$ & $507^{+492}_{-1020}$ & $-36^{+47}_{-0}$ & \nodata & $-65^{+44}_{-0}$ & $59^{+54}_{-46}$ & $18^{+165}_{-195}$ & $609^{+785}_{-1556}$ & $72^{+440}_{-306}$ & $-74^{+616}_{-422}$ & \nodata\\
  \eVS He $\gamma$ & \nodata & $-507^{+106}_{-123}$ & \nodata & $77^{+81}_{-71}$ & \nodata & $-19^{+320}_{-234}$ & $702^{+478}_{-658}$ & $655^{+283}_{-470}$ & \nodata & \nodata\\
  \eVS He $\delta$ & \nodata & \nodata & \nodata & \nodata & \nodata & \nodata & \nodata & \nodata & \nodata & \nodata\\
  \eVS He $\epsilon$ & \nodata & $-93^{+54}_{-59}$ & \nodata & \nodata & \nodata & \nodata & \nodata & \nodata & \nodata & \nodata\\
  \hline
  \eVS He series & $-291\pm({>}1000)$ & $-158^{+35}_{-56}$ & \nodata & $-72^{+25}_{-28}$ & $135^{+205}_{-234}$ & $-86^{+26}_{-41}$ & $47^{+37}_{-52}$ & $2^{+8}_{-12}$ & \nodata & $-21^{+0}_{-379}$ \\
  \hline
 \end{tabular}\\
 }~\\
 {\bfseries Notes.}
 A negative velocity indicates a blue shift, due to the absorbing material moving toward the observer.
 Rows labeled with ``Ly/He series'' show the results from modeling the complete absorption line series
 of the corresponding ion at once with a single physical model (\Sect{sec:absorptionLineSeries}).
\end{table*}
\begin{table*}{\centering
 \caption{Results from the Detected Absorption Lines from H- and He-Like Ions: Column Densities in $10^{16}\,\rm cm^{-2}$}
 \label{tab:H-He-like_columnDensities}
 \begin{tabular}{c@{\quad}ccccccccccc}
  \hline
  \hline
          & O                  & Ne                 & Na                 & Mg                & Al                & Si                & S                 & Ar                & Ca                & Fe          \\
  H-like  &   \sc viii         &    \sc x           &    \sc xi          &    \sc xii        &    \sc xiii       &    \sc xiv        &   \sc xvi         &    \sc xviii      &    \sc xx         &    \sc xxvi \\
  \hline
  \eVS Ly $\alpha$ & $3^{+3}_{-1}$ & $3.5^{+0.2}_{-0.3}$ & \nodata & $5.2^{+0.2}_{-0.3}$ & $1.8\pm0.5$ & $7.1^{+0.3}_{-0.4}$ & $5.0^{+0.6}_{-1.1}$ & $2.2^{+1.2}_{-1.8}$ & $2\pm2$ & \nodata\\
  \eVS Ly $\beta$ & $15^{+3}_{-5}$ & $10.8^{+1.2}_{-1.3}$ & \nodata & $8.0^{+1.6}_{-1.8}$ & $3^{+3}_{-2}$ & $18^{+5}_{-6}$ & \nodata & $7^{+9}_{-7}$ & \nodata & \nodata\\
  \eVS Ly $\gamma$ & $20^{+5}_{-8}$ & $51^{+5}_{-4}$ & \nodata & \nodata & \nodata & $19\pm13$ & $18^{+19}_{-15}$ & \nodata & \nodata & \nodata\\
  \eVS Ly $\delta$ & $11^{+13}_{-20}$ & $82^{+9}_{-7}$ & \nodata & $8^{+18}_{-8}$ & \nodata & \nodata & $29^{+34}_{-29}$ & \nodata & \nodata & \nodata\\
  \hline
  \eVS Ly series & $36^{+18}_{-12}$ & $42^{+10}_{-4}$ & $2.3^{+0.4}_{-0.5}$ & $7.2\pm0.6$ & $1.4\pm0.5$ & $10.1\pm0.8$ & $15^{+12}_{-5}$ & $11^{+1}_{-6}$ & \nodata & \nodata \\
  \hline
  \ \\
  \hline
  \hline
          & O                  & Ne                 & Na                 & Mg                & Al                & Si                & S                 & Ar                & Ca                & Fe         \\
  He-like &   \sc vii          &    \sc ix          &    \sc x           &    \sc xi         &    \sc xii        &    \sc xiii       &   \sc xv          &    \sc xvii       &    \sc xix        &    \sc xxv \\
  \hline
  \eVS He $\alpha$ & \nodata & $1.5\pm0.2$ & \nodata & $1.57^{+0.13}_{-0.15}$ & $0.5^{+0.4}_{-0.1}$ & $1.94^{+0.11}_{-0.18}$ & $2.3^{+0.4}_{-0.5}$ & $1.3\pm0.6$ & $1.0^{+0.9}_{-0.8}$ & $15^{+3}_{-4}$\\
  \eVS He $\beta$ & $4^{+3}_{-6}$ & $2.8^{+0.8}_{-0.4}$ & \nodata & $5.6\pm0.9$ & $3.8^{+1.2}_{-0.6}$ & $9^{+2}_{-3}$ & $16^{+6}_{-9}$ & $4\pm4$ & $11^{+5}_{-7}$ & \nodata\\
  \eVS He $\gamma$ & \nodata & $3.0\pm1.5$ & \nodata & $8\pm2$ & \nodata & $8^{+5}_{-6}$ & $23^{+11}_{-14}$ & $21^{+11}_{-10}$ & \nodata & \nodata\\
  \hline
  \eVS He series & $0^{+4}_{-0}$ & $6.3^{+1.3}_{-1.6}$ & \nodata & $5^{+2}_{-1}$ & $0.8^{+0.2}_{-0.4}$ & $12\pm3$ & $15^{+2}_{-3}$ & $8^{+1}_{-4}$ & \nodata & $146^{+80}_{-0}$ \\
  \hline
 \end{tabular}\\
 }~\\
 {\bfseries Notes.}
 The column densities for the single lines have been calculated using \Eq{eq:COGlin},
 assuming that the line is on the linear part of the curve of growth,
 which underestimates the column density for saturated lines.
 Rows labeled with ``Ly/He series'' show the results from modeling complete absorption line series (\Sect{sec:absorptionLineSeries}).
\end{table*}
The high-resolution spectra reveal a large number of absorption
lines of highly ionized ions.
The 1.5--15\,\AA{} range 
is shown in \Fig{fig:spectrum}
as the ratio between the data and the continuum-model.
As the line profiles are not fully resolved,
we model each line $l$ with a Gaussian profile $G_l$.
In terms of the continuum flux model, $F_\mathrm{cont}$, 
the global model reads
\begin{equation}
 F(\lambda) \;\;=\;\; \mathrm{e}^{-\tau(\lambda)} \;\cdot\; F_\mathrm{cont}(\lambda)
            \;\;=\;\; \left[1+\sum_l G_l(\lambda)\right] \;\cdot\; F_\mathrm{cont}(\lambda) \;\;.
 \label{eq:linemodel}
\end{equation}
From a Gaussian's centroid wavelength $\lambda$ 
and the rest wavelength~$\lambda_0$ of the identified line,
the radial velocity
\begin{equation}
 v \;\;=\;\; (\lambda-\lambda_0)/\lambda_0\cdot c
 \label{eq:v}
\end{equation}
of the corresponding absorber can be inferred.
With the definition in \Eq{eq:linemodel},
the norm of $G_l$ is just the equivalent width:
\begin{equation}
 W_{\lambda,l} \;\;:=\;\; \int\!\! \left[ 1 - F_l(\lambda) / F_\mathrm{cont}(\lambda) \right] \;\mathrm{d}\lambda
               \;\;=\;\; \int\!\! G_l(\lambda) \;\mathrm{d}\lambda
 \label{eq:EW}
\end{equation}
$W_{\lambda,l}$ is related to the absorber's column density,
as a bound--bound transition $i\rightarrow j$
(with the rest frequency $\nu_0$ and \mbox{oscillator} strength $f_{ij}$)
in an absorbing plasma with column density $N_i$
creates the following line profile
(\citealp[see][\mbox{\Section{}9-2}]{Mihalas_StellarAtmospheres}):
\begin{equation}
 \mathrm{e}^{-\tau(\nu)} \;\;=\;\; \exp\left\{ -N_i \: \frac{f_{ij}\,\sqrt{\pi}\,e^2}{m_\mathrm{e}\,c\,\Delta\nu_\mathrm{D}} \; H\!\left({\frac{\Gamma}{4\pi\:\Delta\nu_\mathrm{D}}},\;{\frac{\nu-\nu_0}{\Delta\nu_\mathrm{D}}}\right)\right\}
 \label{eq:absorptionlineprofile}
\end{equation}
Assuming pure radiation damping, the damping constant $\Gamma$ equals the Einstein coefficient $A_{ji}$.
The Doppler broadening $\Delta\nu_\mathrm{D} = \nu_0 \cdot \xi_0/c$
is given by 
${\xi_0}^2 \:=\: {2kT}/{m_\mathrm{ion}} + v_\mathrm{turb}^2$,
i.e., is due to the thermal and turbulent velocities of the plasma.
For optically thin lines (with $\tau(\nu)\ll1$), the equivalent width is
independent of $\Delta\nu_\mathrm{D}$,
such that the absorbing column density can be inferred
(\citealp[][\ifApJ{Equation{\color{red}s}~3{\color{red}--}48}{Eq.~3-48}]{Spitzer1978_ISMprocesses};
 \citealp[][\ifApJ{Section~}{\S}10-3]{Mihalas_StellarAtmospheres}):
\begin{equation}
 N_i \;=\; \frac{m\,c^2\cdot W_\lambda}{\pi\,e^2\cdot f_{ij}\,\lambda_0^2} \;\;=\;\;
           \frac{1.13\!\times\!10^{17}\:\mathrm{cm}^{-2}}{f_{ij}} \cdot \left(\frac{\lambda_0}{\mbox{\AA}}\right)^{\!\!\!\!-2} \!\!\!\!\cdot \left(\frac{W_\lambda}{\mbox{m\AA}}\right)
\label{eq:COGlin}
\end{equation}
If the lines are, however, saturated, $W_\lambda$ depends on $\Delta\nu_\mathrm{D}$ as well,
and one has to construct the full ``curve of growth'' with several lines from a common ground state $i$
in order to constrain~$N_i$ \citep[see, e.g.,][]{Kotani2000}.

We have therefore performed a systematic analysis of absorption line series:
H-like ions are detected by their $1\mathrm{s}\rightarrow n\,\mathrm{p}$ Lyman series
and He-like ions by their $1\mathrm{s}^2\rightarrow1\mathrm{s}\:n\,\mathrm{p}$ resonance absorption series,
see Table~\ref{tab:H-He-like}.
For those lines that are clearly detected and not obviously affected by blends,
the measured velocity shifts (\Equation~\ref{eq:v}) are shown in Table~\ref{tab:H-He-like_velocityShifts}.
Most of the lines are detected at rather low projected velocity ($|v|<$200\,km\,s$^{-1}$).
Note that the systemic velocity of Cyg\,X-1/HDE\,226868 is $(-7.0\pm0.5)$\,km\,s$^{-1}$ \citep{Gies2003},
and that the radial velocity of both the supergiant and the black hole vanishes at orbital phase $\phi_\mathrm{orb}=0$,
while the radial component of the focused stream should be maximal
(likely to be up to 720\,km\,s$^{-1}$, see \Sect{sec:winddensity}).
The column densities in Table~\ref{tab:H-He-like_columnDensities} are calculated using \Eq{eq:COGlin},
assuming that the line is on the linear part of the curve of growth.
As the strongest lines are, however, often saturated,
\Eq{eq:COGlin} predicts too small column densities from them.
For weak lines, however, the equivalent width is most likely to be
overestimated such that the quoted values may rather be upper limits.
The properties of the lines (Einstein $A$-coefficients and quantum multiplicities,
which determine the oscillator strength, as well as the rest wavelengths)
are taken from \citet{Verner1996} and
\textsc{atomdb}\footnote{See \url{http://cxc.harvard.edu/atomdb/}\ifApJ{}{.}} version~1.3.1.

\subsection{Line Series of H-/He-Like and Fe L-Shell Ions} \label{sec:absorptionLineSeries}
As an alternative to the curve of growth,
we chose to develop a model which implements the expected line profiles (\Equation~\ref{eq:absorptionlineprofile}) directly
for all transitions of a series from a common ground state $i$.
The model contains $N_i$, $\xi_0$, and the systemic shift velocity $v$ (\Equation~\ref{eq:v})
as fit parameters, and thus avoids the use of equivalent widths at all.
This approach allows for a systematic treatment of the iron L-shell transitions as well,
which are often blended with other lines
such that the different contributions to a line's equivalent width can hardly be separated
when only single Gaussians are used.
As an example, \Fig{fig:FeXIX} shows the \Ion{Fe}{xix} complex between 12.8\,\AA{} and 14\,\AA{}.
Lines from different \Ion{Fe}{xix} transitions overlap,
and so does the strong \Ion{Ne}{ix} r-line.
Furthermore, the absorption features are often rather weak
and no prominent lines can be fitted, whereas the line series model can still be applied.
Although a disadvantage of this approach is the larger computational effort,
it usually allows us to constrain the parameters of a line series more tightly.

The model with physical absorption line series fits the data hardly worse
than the model with single Gaussian lines:
$\chi^2$ of 12812 instead of 12180 before (see Table~\ref{tab:continuum}) is obtained.
The results are presented in the last row for the whole Ly/He series
of Tables~\ref{tab:H-He-like_velocityShifts} and \ref{tab:H-He-like_columnDensities}
for the H-/He-like ions,
and in Table~\ref{tab:lineSeries} for the Fe L-shell ions.
The column densities inferred from the series model are generally in good agreement 
with the values derived from the single Gaussian fits
for the higher transitions of H- or He-like ions (Table~\ref{tab:H-He-like_columnDensities}),
while the $\alpha$\ifApJ{---}{ -- }sometimes even $\beta$\ifApJ{---}{ -- }lines are saturated.
These measurements can be used as input for wind or photoionization models,
although a few columns are rather badly constrained
if the thermal velocity $\xi_0$ is left as a free parameter.
As the line profiles are not resolved, there is a degeneracy
between Doppler-broadened lines and narrow, but saturated lines.
The notable fact that no strong wavelength shifts are observed
is, however, independent of this degeneracy: almost all line series are consistent
with a velocity between $-200$\,km\,s$^{-1}$ and +200\,km\,s$^{-1}$.
Lower ionization stages of the same element are usually seen at higher blueshift,
like in the sequence \Ion{Fe}{xxiii}--\textsc{xxii}--\textsc{xxi}--\textsc{xx}.
\begin{figure}\centering
 \includegraphics[width=0.95\columnwidth]{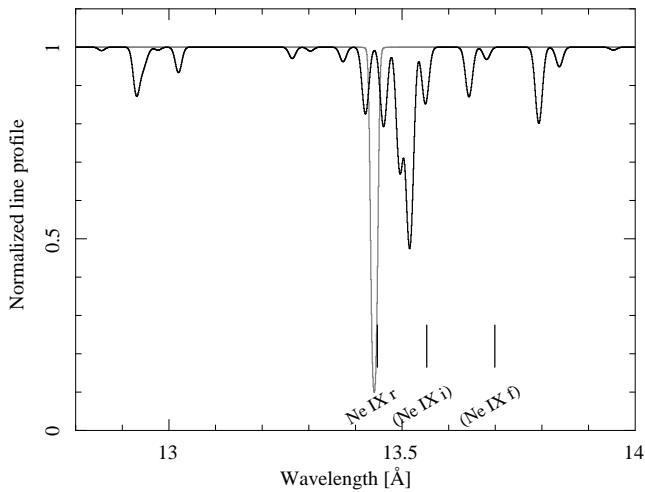}
 \caption{Absorption line series of \Ion{Fe}{xix} between 12.8\,\AA{} and 14\,\AA.
          The \Ion{Ne}{ix} r line which blends with the complex at 13.4--13.6\,\AA{} is also shown,
          and the expected positions of the \Ion{Ne}{ix} i and f emission lines are indicated.
         }
 \label{fig:FeXIX}
\end{figure}
\begin{table}\centering
 \caption{Parameters of the Absorption Line Series for Fe L-Shell Ions}
 \label{tab:lineSeries}
 \begin{tabular}{cccc}
  \hline
  \hline
  Ion & $v$                & $N(\mathrm{ion})$             & $\xi_0$ \\
      & $(\rm km\,s^{-1})$ & $(10^{16}\,\mathrm{cm}^{-2})$ & $(\rm km\,s^{-1})$\\
  \hline
  \eVS Fe\,\textsc{xxiv} & $21^{+31}_{-30}$ & $4.9^{+0.7}_{-1.0}$ & $82^{+21}_{-22}$ \\
  \eVS Fe\,\textsc{xxiii} & $120^{+68}_{-94}$ & $2.2^{+0.4}_{-0.2}$ & $410^{+68}_{-140}$ \\
  \eVS Fe\,\textsc{xxii} & $-27^{+33}_{-35}$ & $2.3^{+0.2}_{-0.4}$ & $193^{+61}_{-64}$ \\
  \eVS Fe\,\textsc{xxi} & $-156^{+30}_{-29}$ & $2.5^{+0.2}_{-0.4}$ & $260^{+59}_{-68}$ \\
  \eVS Fe\,\textsc{xx} & $-283^{+33}_{-48}$ & $3.0^{+0.3}_{-0.2}$ & $333^{+57}_{-92}$ \\
  \eVS Fe\,\textsc{xix} & $-116^{+46}_{-41}$ & $3.1^{+0.2}_{-0.3}$ & $251^{+80}_{-68}$ \\
  \eVS Fe\,\textsc{xviii} & $-60^{+51}_{-54}$ & $1.8^{+0.3}_{-0.2}$ & $203^{+72}_{-56}$ \\
  \eVS Fe\,\textsc{xvii} & $-277^{+56}_{-42}$ & $1.4\pm0.3$ & $78^{+26}_{-25}$ \\
  \hline
 \end{tabular}
\end{table}

\begin{figure}\centering
 \includegraphics[width=0.95\columnwidth]{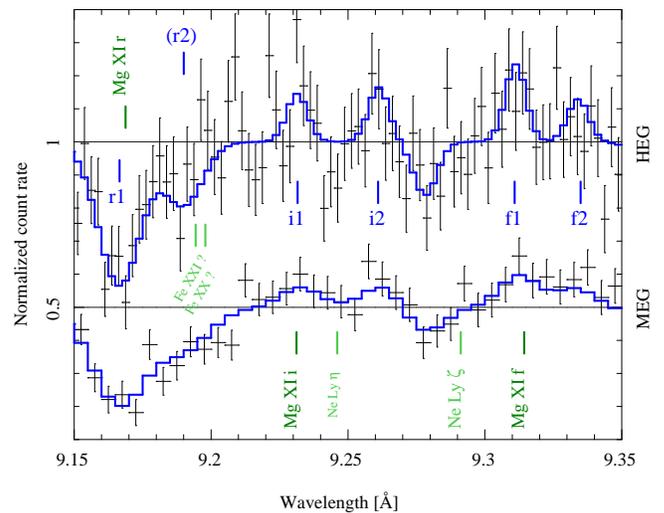}
 \caption{Mg\,\textsc{xi} triplet of resonance (r), intercombination (i), and forbidden (f) lines
          in the \Chandra{}-HETGS spectrum.
          The MEG data are shown with an offset of $-0.5$.
          While the ion/line labels indicate the rest wavelengths,
          the abbreviations r$n$, i$n$, and f$n$ label the actual line positions (see Table~\ref{tab:MgXIlines}).}
 \label{fig:MgTriplet}
\end{figure}
\begin{table}{\centering
 \caption{Parameters of the Lines of the Mg\,\textsc{xi} Triplet}
 \label{tab:MgXIlines}
 \begin{tabular}{cccccc}
  \hline
  \hline
  Line                   & Component & $\lambda$                    & $v$               & $W_\lambda$\\
                         &           & (\AA)                        & (km\,s$^{-1}$)    & (m\AA)\\
  \hline
  \eVS Mg\,\textsc{xi} r~& r1        & $9.167^{+0.001}_{-0.002}$ & $-70^{+35}_{-49}$    & $-8.6^{+0.8}_{-0.7}$ \\
  \eVS Mg\,\textsc{xi} r?& (r2)      & $9.190^{+0.004}_{-0.003}$ & $\left(700^{+120}_{-90}\right)$ & $-3.5\pm0.9$ \\
  \hline
  \eVS Mg\,\textsc{xi} i~& i1        & $9.232\pm0.003$           & $14^{+107}_{-105}$   & $+1.5^{+1.5}_{-0.6}$\\
  \eVS Mg\,\textsc{xi} i~& i2        & $9.261^{+0.002}_{-0.003}$ & $968^{+52}_{-110}$   & $+1.7^{+1.1}_{-0.9}$\\
  \hline
  \eVS Mg\,\textsc{xi} f~& f1        & $9.311^{+0.004}_{-0.001}$ & $-110^{+130}_{-34}$  & $+2.5^{+0.9}_{-1.2}$\\
  \eVS Mg\,\textsc{xi} f~& f2        & $9.328^{+0.009}_{-0.003}$ & $454^{+293}_{-111}$  & $+1.3^{+0.9}_{-1.0}$ \\
  \hline
 \end{tabular}\\}~\\
 {\bfseries Notes.}
 The shift velocity $v$ and equivalent width $W_\lambda$ are given by
 \Eqs{eq:v} and (\ref{eq:EW}).
 Note that the (r2) component is likely to be caused by absorption from blueshifted
 transitions of Fe\,\textsc{xxi}, Fe\,\textsc{xx}, and Ne\,\textsc{x}.
\end{table}
\subsection{Emission Lines from He-Like Ions} \label{sec:HeTriplets}
The transitions between the 1s$^2$ ground state and the 1s2s or 1s2p excited states of He-like ions
lead to the triplet of forbidden (f), intercombination (i), and resonance (r) line, see Table~\ref{tab:H-He-like}.
These lines provide a density and temperature diagnostics of an \emph{emitting} plasma via the ratios 
$ R(n_\mathrm{e}) \::=\: F($f$)\,/\,F$(i) and $G(T_\mathrm{e}) \::=\: [F($i$)+F($f$)]\,/\,F$(r)
of the fluxes $F$ in the r-, i-, and f-line
\citep[see, e.g.,][]{GabrielJordan1969,PorquetDubau2000}.
In this observation, the dipole-allowed resonance transitions are seen as absorption lines,
as an \emph{absorbing} plasma is detected in front of the X-ray source.
For the same reason, we cannot use the Fe L-shell density diagnostics \citep{Mauche2005}
as, e.g., the \ion{Fe}{22} emission lines at 11.77\,\AA{} and 11.92\,\AA{}
used by \citet{Mauche2003} are both seen in absorption.
But we can still use the detected He-i and -f emission lines
to estimate the density via the $R$-ratio,
noting the caveat that the densities are systematically overestimated
in the presence of an external UV radiation field,
as photoexcitation of the $\mathrm{1s2s\:\left(^3S_1\right)\rightarrow1s2p\:\left(^3P\right)}$ transition
depopulates the upper level of the f-line in favor of the i-line
and leads to a lower $R$-ratio \citep[see, e.g.,][]{MeweSchrijver1978,Kahn2001,Wojdowski2008}.

The i- and f-lines of the \Ion{Mg}{xi} triplet are seen as two distinct components each
-- one with almost no shift, which is consistent with the r-absorption line,
   and the other one at a redshift of 400--1000\,km\,s$^{-1}$
(see \Fig{fig:MgTriplet} and Table~\ref{tab:MgXIlines}).
Given these two emission components and their Doppler shifts,
one could be tempted to identify the absorption feature at 9.19\,\AA{}
with a second redshifted Mg\,\textsc{xi}\,r line,
but our model for the complete absorption line series (\Sect{sec:absorptionLineSeries})
predicts that the blueshifted $\mathrm{2s\rightarrow4p}$ transitions from the ground states
of Fe\,\textsc{xxi} and Fe\,\textsc{xx}, as well as the Ne\,\textsc{x}~$\mathrm{1s\rightarrow10p}$ transition,
account for most of the absorption seen at 9.19\,\AA.
Table~\ref{tab:MgXIR} shows the $R$-ratios obtained for the two pairs of lines
as well as the corresponding densities
according to the calculations of \citet[Fig.~8e]{PorquetDubau2000},
\emph{neglecting the influence of the UV radiation of HDE\,226868 on the $R$-ratio}.
The unshifted lines would then be caused by a plasma
with an electron density $n_\mathrm{e,1}\approx5\!\times\!10^{12}\,$cm$^{-3}$;
the redshifted pair of lines seems to stem from another plasma component
with $n_\mathrm{e,2} \gg n_\mathrm{e,1}$.
A more detailed discussion of these densities is presented in \Sect{sec:winddensity}.
\begin{table}{\centering
 \caption{Electron Densities Corresponding to $R=F(\mathrm{f})/F(\mathrm{i})$ Ratios}
 \label{tab:MgXIR}
 \begin{tabular}{cccc}
  \hline
  \hline
  & & \multicolumn{2}{c}{Model Calculations by} \\
  & & \multicolumn{2}{c}{\citet{PorquetDubau2000}$^a$} \\
  \cline{3-4}
  & & $R$(Mg\,\textsc{xi}) & $n_\mathrm{e}$ \\
  Component & Measurement & & ($10^{12}\,$cm$^{-3})$ \\
  \hline
  First pair       &  & 2.3 & 1.0\dots  4.5 \\
  of i and f lines & $R_1($Mg\,\textsc{xi}$)=1.6^{+0.7}_{-0.6}$ \large$\Big\{$ 
                      & 1.6 & 3.5\dots 10 \\
  (unshifted)      &  & 1.0 & 8  \dots 22 \\
  \hline
  Second pair      &  & 1.3 &  5\dots 15 \\
  of i and f lines & $R_2($Mg\,\textsc{xi}$)=0.8\pm0.5$ \large$\Big\{$
                      & 0.8 & 12\dots 28 \\
  (redshifted)     &  & 0.3 & 43\dots 91 \\
  \hline
 \end{tabular}\\
 }~\\
 {\bfseries Note.}
 $^a$ A temperature between 0.3 and 8 MK is assumed\\
 and \emph{UV-photoexcitation is not considered}.
\end{table}

The S/N of the spectrum above 20\,\AA{} is not high enough to describe the \Ion{O}{vii} triplet.
The \Ion{Ne}{ix} triplet blends with several \Ion{Fe}{xix} lines (\Fig{fig:FeXIX}).
\Ion{Na}{x} i- and f-lines are likely to be present with several components,
but those cannot be distinguished clearly.
Similar as for Mg, there are also two i-lines of Al\,\textsc{xii} at shifts of $-477^{+120}_{-240}$\,km\,s$^{-1}$
and $+333^{+156}_{-228}$\,km\,s$^{-1}$, respectively,
but no f-line is detected due to a blend with the strong Mg\,\textsc{xi} He\,$\beta$ absorption line
and especially an Fe\,\textsc{xxii} 1s$^2$2s2p(2s$\rightarrow$5p) absorption line at 7.87\,\AA.
Similarly, the Si\,\textsc{xiii} i-line is not detected as it overlaps with absorption features
probably due to nearly neutral Si K-edge structures.
(Furthermore, the flux of the Si\,\textsc{xiii} f-line might be underestimated
 as it blends with the Mg\,\textsc{xii} Ly\,$\gamma$ line.)
The i- and f-lines of He-like sulfur are detected with $R($S\,\textsc{xv}$)=0.61^{+1.31}_{-0.60}$,
but S\,\textsc{xv} has not been modeled by \citet{PorquetDubau2000}.
Below 4\,\AA{}, the S/N below is again not high enough
to resolve further triplets from heavier elements such as argon, calcium, or iron.

\section{Discussion And Conclusions}\label{sec:discuss}
In the following, we discuss our results and derive constraints on the
stellar wind in the accretion region.

\subsection{Velocity and Density of the Stellar Wind}\label{sec:winddensity}
While a spherically symmetric model for the stellar wind in the HDE\,226868/\Cyg{} system
can be excluded by observations \cite[see, e.g.,][]{GiesBolton1986_III,Gies2003,Miller2005,Gies2008},
a symmetric velocity law 
\begin{equation}
 v(r) \;\;=\;\; v_\infty \;\cdot\; f(r/R_\star)
 \label{eq:vr}
\end{equation}
is usually assumed to obtain a first estimate of the particle density in the wind.
The fraction $f$ of the terminal velocity $v_\infty$ is often parameterized by
\citep[\Equation~3]{LamersLeitherer1993}
\begin{equation}
 f(x) \;\;=\;\; f_0 \:\;+\:\; \left(1-f_0\right) \cdot \left(1-1/x\right)^\beta
 \quad (\mbox{for } x\ge1)
 \label{eq:f}
\end{equation}
with $f_0:=v_0/v_\infty$, where $v_0$ is the velocity at the base of the wind ($x=1$).
The simple model for the radiatively driven wind of isolated stars by \citet{CastorAbbottKlein1975}
is obtained for $\beta=1/2$.
The photoionization of the wind, however,
suppresses its acceleration \citep{Blondin1994},
such that a smaller $f$ (e.g., a larger $\beta$ within the same model)
is required in the Str\"omgren zone.
\citet{GiesBolton1986_III} have explained the orbital variation
of the 4686\,\AA{} He\,\textsc{ii} emission line profile of \Cyg{}
with a similar model for the focused wind,
where $v_\infty$, $R_\star$, and $\beta$
depend on the angle~$\theta$ from the binary axis.
$\beta$~was interpolated between 1.60 and 1.05
for $\theta$ between $0$ and $20^\circ$.
The value $\beta(\theta\!=\!20^\circ)=1.05$ is, however,
often used for a spherically symmetrical wind as well
\citep{Lachowicz2006,Szostek2007,Poutanen2008}.
\citet{Vrtilek2008} use $f_0=0.01$ to avoid numerical singularities
and fit the (relatively low) value $\beta\approx0.75$
to their models for UV lines.
$v_0$ is likely to be of the order of the thermal velocity of H atoms,
which is $(2kT/m_\mathrm{H})^{1/2}=23\,$km\,s$^{-1}$
and corresponds to $f_0=0.011$
for HDE\,226868's effective temperature $T_\mathrm{eff}=32$\ifApJ{,}{\,}000\,K 
and $v_\infty=2100\,$km\,s$^{-1}$ \citep{Herrero1995}.

\begin{table}\centering
 \caption{Solutions to \Eq{eq:nr_numbers} for the Model of \Eq{eq:f}}
 \label{tab:density_solutions}
 \begin{tabular}{cccccc}
  \hline
  \hline
  $n_\mathrm{H}(x)$      & $d(x)$ &  $f_0$  & $\beta$  & $x$      & $2100 \cdot f(x)$ \\
  (10$^{10}$\,cm$^{-3}$) &        & $=v_0/v_\infty$  & & $=r/R_\star$ & for $\beta=\beta_\mathrm{max}$ \\
  \hline \eVS
  440                    & 200  & 0.005 & $<\infty$ & 1 & 11\\
  \hline \eVS
  110                    &  50  & 0.011 & $\le1$    & $\le1.01$ & $\le41$\\
  110                    &  50  & 0.011 & $\le2$    & $\le1.08$ & $\le36$\\
  110                    &  50  & 0.011 & $\le3$    & $\le1.18$ & $\le30$\\
  \hline \eVS
  22                     &  10  & 0.011 & $\le1$    & $\le1.08$ & $\le180$\\
  22                     &  10  & 0.011 & $\le2$    & $\le1.29$ & $\le127$\\
  22                     &  10  & 0.011 & $\le3$    & $\le1.48$ & $\le\phantom{1}95$\\
  \hline
  4.4                    &   2  & 0.011 & $\le1$    & $\le1.36$ & $\le570$\\
  4.4                    &   2  & 0.011 & $\le2$    & $\le1.69$ & $\le368$\\
  4.4                    &   2  & 0.011 & $\le3$    & $\le1.97$ & $\le271$\\
  \hline
 \end{tabular}\\
 {\bfseries Note.} \mbox{The binary separation $a=41\,R_\odot$ corresponds to $x_a=a/R_\star=2.4$.}
\end{table}
With the mean molecular weight per H atom, $\mu\approx1.4$,\footnote{%
 The helium abundance per H atom is $\approx$10\% \citep{Wilms2000}.
 \label{footnote:HeAbund}} the~continuity equation
$\dot M_\star \;=\; \mu\, m_\mathrm{H}\, n_\mathrm{H}(r) \:\cdot\: 4\pi r^2\, v(r)$
gives the following estimate for the hydrogen density profile:
\begin{equation}
 n_\mathrm{H}(r) \;=\; \frac{\dot M_\star/(\mu\,m_\mathrm{H})}{4\pi R_\star^2 v_\infty} \:\cdot d(r/R_\star)
\;\;\mathrm{with}\;\; d(x) \;=\; x^{-2}/f(x)
 \label{eq:nr}
\end{equation}
Using the parameters of HDE\,226868
($\dot M_\star=3\!\times\!10^{-6}\,M_\odot\,\mathrm{yr}^{-1}$, $R_\star=17\,R_\odot$;
 \citealp{Herrero1995}; also summarized by \citealt{Nowak1999_II}, Table~1),
\Eq{eq:nr} predicts
\begin{equation}
 n_\mathrm{H}(r) \;\;=\;\; 2.2\!\times\!10^{10}\,\mathrm{cm}^{-3} \;\cdot\; d(r/R_\star) \;\;.
 \label{eq:nr_numbers}
\end{equation}
\Eq{eq:nr} can only be solved within the model of \Eq{eq:f}%
\mbox{\ifApJ{---}{ --~}}or~any other model
for $f$ in \Eq{eq:vr} with $f(x)\ge f_0$\ifApJ{---}{ -- }for
\begin{equation}
 d \;\;<\;\; f_0^{-1} \quad\quad\mbox{and}\quad\quad 1 \;\;\le\;\; x \;\;\le\;\; \sqrt{1/(f_0d)\;} \;\;.
 \label{eq:density_solution}
\end{equation}
For $f_0\approx0.01$, the value $d\approx190$\ifApJ{---}{ -- }%
which would be required to explain the density\footnote{%
 A plasma of fully ionized H and He  contains $\approx$1.2 electrons per H atom.
} $n_\mathrm{H,1} \approx 4.2\!\times\!10^{12}\,$cm$^{-3}$
obtained from the unshifted Mg\,\textsc{xi} triplet (\Sect{sec:HeTriplets})%
\ifApJ{---}{ -- } can never be reached within our model for the continuous spherical wind.
This result shows that the density is likely to be overestimated by an $R$-ratio analysis
which ignores the strong UV-flux of the O9.7 star.
Table~\ref{tab:density_solutions} lists therefore some solutions
to \Eq{eq:nr_numbers} for much lower densities as well.
Due to the additional constraint
that the radial velocity is less than 100\,km\,s$^{-1}$ (Table~\ref{tab:MgXIlines}),
we suggest that the first emission component stems from close to the stellar surface.
Although the simple wind model of \Eq{eq:f}
may not be appropriate in this region,
the results of Table~\ref{tab:density_solutions} are rather insensitive
to the assumptions on the velocity law, as a wide range of $\beta$ values was considered,
but mostly depend on the wind's initial velocity $v_0$,
for which we have used a reasonable estimate.
Note that $f_0$ and $d$ also depend on $v_\infty$,
but their product, which is important in \Eq{eq:density_solution}, does not.

In spite of the systematic errors of the (absolute) density analysis with the $R$-ratio,
we infer that the second plasma component is much denser relative to the first one.
As it is seen at a larger redshift of 400--1000\,km\,s$^{-1}$,
we favor its identification with the focused wind.
The two emission components could also be caused coincidentally by dense clumps in the stellar wind
\citep[which are common for O-stars; e.g.,][]{Oskinova2007,LepineMoffat2008},
but the interpretation as a slow base of the wind close to the stellar surface
and a focused wind between the accreting black hole and its donor star
provides a consistent description of both emission components:
an undisturbed wind (with $v_\infty$ and $f_0$ as above, and $\beta=1.05\rightarrow1.6$)
would reach a velocity of $v(a)\approx900\leftarrow1200$\,km\,s$^{-1}$ in the distance of the black hole.
The focused wind in the orbital plane would then be detected
with a projected velocity $v(a)\sin i=500\leftarrow700$\,km\,s$^{-1}$ at $\phi_\mathrm{orb}=0$.
While photoionization reduces the efficiency of acceleration for the spherical wind,
the denser focused wind is less strongly affected due to self-shielding
and reaches the expected velocity.
It is an observational fact that the focused wind has another ionization structure:
optical emission lines (H$\alpha$, He\,\textsc{ii} $\lambda4686$)
from ions which only exist at a low ionization parameter
have been observed in the focused stream \citep[see, e.g.,][]{GiesBolton1986_III,Ninkov1987b,Gies2003}.
\mbox{For the spherical wind,} however, \citet{Gies2008} have conjectured that the part
between \Cyg{} and the donor star might be completely ionized in the soft state.
Our observations show that the situation in the hard state may be similar.

\subsection{Modeling of the Photoionization Zone} \label{sec:XSTAR}
We use the photoionization code \textsc{XSTAR}~2.1ln7b\footnote{%
 See \url{http://heasarc.gsfc.nasa.gov/lheasoft/xstar/xstar.html}\ifApJ{}{.}} \citep{Kallman2001}
to model the photoionization zone.
As the latter is quite complex (due to the inhomogeneous wind density,
which is strongly entangled with the X-ray flux),
we do not claim to describe the photoionized wind self-consistently,
but only want to derive a first approximation.
For an optically thin plasma,
the relative population of a given atom's ions 
is merely a function of the ionization parameters
\begin{equation}
 \xi(r) \;\;=\;\; \frac{L}{n_\mathrm{H}\,r^2} \;\;=\;\;  \frac{L_{37}}{n_{13}\,r_{12}^2} \;\;\mathrm{erg}\;\mathrm{cm}\;\mathrm{s}^{-1} \;\;,
 \label{eq:ionizationParameter}
\end{equation}
where $L_{37}$ is the ionizing source luminosity above 13.6\,eV
in~$10^{37}\,\mathrm{erg}\;\mathrm{s}^{-1}$,
$n_{13}$ is the hydrogen density in $10^{13}\,\mathrm{cm}^{-3}$
and $r_{12}$~is the distance from the source in $10^{12}\,\mathrm{cm}$.
\begin{figure}\centering
 \includegraphics[width=0.95\columnwidth]{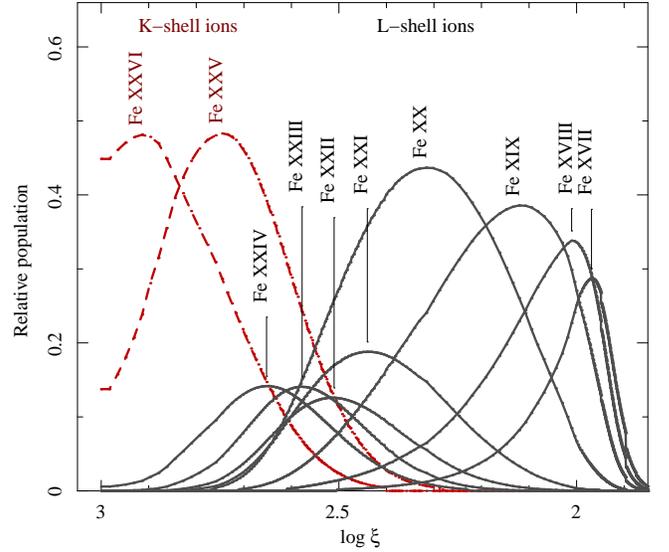}
 \caption{Relative population of the ionization stages of iron as a function of the ionization parameter $\xi=L/(n_\mathrm{H}\,r^2)$,
          as calculated with \textsc{XSTAR} for $L_{37}=3.5$ and $n_{13}=0.01$.}
 \label{fig:IonizationBalance}
\end{figure}
\begin{table}{\centering
 \ifApJ{}{\vskip-5mm}
 \caption{Ionization Parameters for Peak Ion Populations}
 \label{tab:ionParameter}
 \begin{tabular}{cccc}
  \hline
  \hline
  Ion & log$\:\xi$ & FWHM(log$\:\xi$) & \eVS $r_{12}=\left(L_{37}/n_{13}/10^{\log\xi}\right)^{1/2}$\\
  \hline
   Fe\,\textsc{xxvi}  &  2.91  &  0.27  &  0.6--0.8 \\
    Fe\,\textsc{xxv}  &  2.75  &  0.32  &  0.7--1.0 \\
   Fe\,\textsc{xxiv}  &  2.65  &  0.30  &  0.8--1.1 \\
  Fe\,\textsc{xxiii}  &  2.58  &  0.28  &  0.8--1.1 \\
   Fe\,\textsc{xxii}  &  2.51  &  0.31  &  0.9--1.3 \\
    Fe\,\textsc{xxi}  &  2.44  &  0.36  &  0.9--1.4 \\
     Fe\,\textsc{xx}  &  2.32  &  0.44  &  1.0--1.7 \\
    Fe\,\textsc{xix}  &  2.12  &  0.40  &  1.2--1.9 \\
  Fe\,\textsc{xviii}  &  2.01  &  0.23  &  1.6--2.0 \\
   Fe\,\textsc{xvii}  &  1.97  &  0.14  &  1.8--2.1 \\
  \hline
     Ca\,\textsc{xx}  &  2.69  &  0.56  &  0.7--1.2 \\
    Ca\,\textsc{xix}  &  2.39  &  0.54  &  0.9--1.6 \\
  \hline
  Ar\,\textsc{xviii}  &  2.55  &  0.65  &  0.7--1.5 \\
   Ar\,\textsc{xvii}  &  2.22  &  0.50  &  1.0--1.8 \\
  \hline
     S\,\textsc{xvi}  &  2.38  &  0.69  &  0.8--1.7 \\
      S\,\textsc{xv}  &  2.08  &  0.43  &  1.2--2.0 \\
  \hline
    Si\,\textsc{xiv}  &  2.21  &  0.65  &  0.9--2.0 \\
   Si\,\textsc{xiii}  &  1.96  &  0.39  &  1.5--2.4 \\
     Si\,\textsc{iv}  &  1.09  &  0.18  &  4.7--5.7 \\
  \hline
    Mg\,\textsc{xii}  &  2.06  &  0.58  &  1.2--2.3 \\
     Mg\,\textsc{xi}  &  1.82  &  0.31  &  1.8--2.6 \\
  \hline
      Ne\,\textsc{x}  &  1.94  &  0.40  &  1.6--2.5 \\
     Ne\,\textsc{ix}  &  1.74  &  0.24  &  2.1--2.8 \\
  \hline
    O\,\textsc{viii}  &  1.80  &  0.26  &  1.9--2.6 \\
     O\,\textsc{vii}  &  1.62  &  0.22  &  2.5--3.2 \\
  \hline
       N\,\textsc{v}  &  1.51  &  0.01  &  3.25--3.30 \\
  \hline
      C\,\textsc{iv}  &  1.48  &  0.19  &  3.3--4.1 \\
  \hline
 \end{tabular}\\}
 ~\\
 {\bfseries Notes.}
 We list the range in ionization parameter and distance for the maximum ion population (\Fig{fig:IonizationBalance})
 obtained in an \textsc{XSTAR} simulation with $L_{37}=3.5$ and $n_{13}=0.01$.
 Note that the binary separation in units of $10^{12}\,$cm is $a_{12}=2.9$.
\end{table}

For \Cyg{} as observed in the hard state,
extrapolation of the unabsorbed model obtained by our analysis (Table~\ref{tab:continuum})
gives $L_{37}\approx3.5$.
Although there are obviously strong variations in the density,
the \textsc{XSTAR} calculation has been performed with constant $n_{13}=0.01$,
which is the average of $n_\mathrm{H}(r)$ from \Eq{eq:nr_numbers} for $R_\star\le r\le a$.
\Fig{fig:IonizationBalance} shows the resulting population of Fe ions.
The ionization parameters at which the population distributions of some ions of interest peak,
as well as the corresponding full width at half maximum (FWHM),
are presented in Table~\ref{tab:ionParameter}.
We have also included Si\,\textsc{iv}, N\,\textsc{v}, and C\,\textsc{iv}
for a comparison with the work of \citet{Vrtilek2008} and \citet{Gies2008}.
The H- and He-like ions detected with \Chandra{}
only appear at considerable distance from the X-ray source.
Taking into account
that the photons emanating from \Cyg{} first propagate through a lower density wind
and that the (eventually stalled) wind of higher density is only reached in the vicinity of the star,
the actual distances will be even larger than the $r_{12}$ values quoted in Table~\ref{tab:ionParameter}.

\subsection{Origin of Redshifted X-Ray Absorption Lines}
\begin{figure}\centering
 \includegraphics[width=0.95\columnwidth]{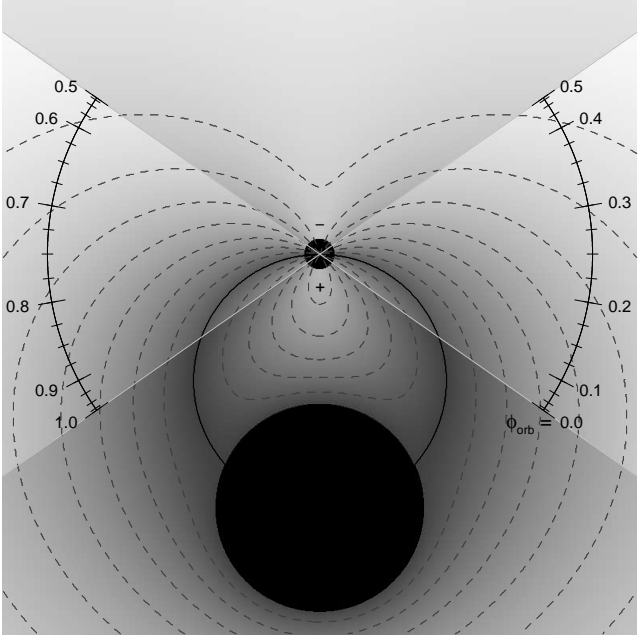}
 \caption{Gray-scale image of $v_\mathrm{rad,BH}(\vecr)$ of \Eq{eq:proj_velocity},
          i.e., projected wind velocity against the black hole.
          A spherically symmetric velocity law given by \Eqs{eq:vr} and (\ref{eq:f})
          with $v_\infty=2100\,\mathrm{km\,s}^{-1}$, $f_0=0.01$, and $\beta=1.05$ is assumed.
          The star and the black hole are shown as filled black circles;
          the size of the latter is the accretion radius $2GM_\mathrm{BH}/v(a)^2$.
          On the black circle passing through the center of the star and through the black hole
          $v_\mathrm{rad,BH}(\vecr)$ is $0$ since $\alpha(\vecr)=90^\circ$.
          \emph{\mbox{Positive} velocities} (+) are seen only \emph{within this circle};
          a lighter gray means a larger redshift.
          \emph{Negative velocities} ($-$) can likewise only occur \emph{outside of this circle};
          a lighter gray here means a larger blueshift.
          \ifApJ{Lines of {\color{red}the}}{The dashed gray lines of} constant \ifApJ{$v$}{$v_\mathrm{rad,BH}$}
          are shown from $-1400\,$km\,s$^{-1}$ to $+1000$\,km\,s$^{-1}$ in steps of 200\,km\,s$^{-1}$.
          The two highlighted sectors contain all observable lines of sight toward the black hole
          (after rotation in this plane) for the inclination $i=35^\circ$.
          The labels show the corresponding orbital phases.
         }
 \label{fig:proj_velocity}
\end{figure}
Additional constraints on the accretion flow can be derived
by considering the Doppler shifts observed in absorption lines
(Sections~\ref{sec:absorptionLines} and \ref{sec:absorptionLineSeries}).
Models for the wind velocity like that of \Eqs{eq:vr} and (\ref{eq:f})
predict a velocity $\vecv(\vecr)$ at the position $\vecr$ in the stellar wind.
But only the projection of $\vecv$ against the black hole, $v_\mathrm{rad,BH}$,
can eventually be observed as radial velocity in an X-ray absorption line.
With the angle $\alpha(\vecr)$ between $\vecv(\vecr)$
and the direction from $\vecr$ toward the black hole, $v_\mathrm{rad,BH}(\vecr)$ is
\begin{equation}
 v_\mathrm{rad,BH}(\vecr) \;\;=\;\; \cos\alpha(\vecr) \;\cdot\; |\vecv(\vecr)| \;\;.
 \label{eq:proj_velocity}
\end{equation}
Assuming a velocity field with radial symmetry with respect to the star,
$\alpha(\vecr)$ is just the angle at $\vecr$ between the star and the black hole.
For example, $\alpha$ is $90^\circ$ (and $v_\mathrm{rad,BH}$ is therefore 0)
on the sphere containing the center of the star and the black hole diametrically opposed.
Redshifted absorption lines can only be observed from wind material inside this sphere,
while the part of the wind outside of it is always seen at a blueshift
-- a fact which is independent of the assumed velocity field
as long as it is directed radially away from the star.

For a spherically symmetrical wind model,
any line of sight can be rotated to an equivalent one
in a half-plane limited by the binary axis.
Only a sector with half-opening angle~$i$ can be observed
unless the system's inclination is $i=90^\circ$.
\Fig{fig:proj_velocity} shows the projected velocity $v_\mathrm{rad,BH}(\vecr)$
for a simple wind model and two sectors containing all observable lines of sight
toward \Cyg{} for an inclination of $i=35^\circ$.

We now investigate the region where the projected wind velocity (\Fig{fig:proj_velocity})
is compatible with the observed Doppler shifts
(Tables~\ref{tab:H-He-like_velocityShifts} and \ref{tab:lineSeries}).
We are confident that this method allows for the identification of the absorption regions,
as the low observed velocities are always found close to the $\alpha=90^\circ$ sphere
-- independent of the wind model.
For most of the investigated ions (\Ion{Ne}{x}, \Ion{Mg}{xii}, \Ion{Si}{xiv} and \textsc{xiii},
\Ion{S}{xv}, \Ion{Ar}{xvii}, \Ion{Fe}{xix} and \textsc{xviii})
the inferred distance from the black hole
agrees with the predictions of Table~\ref{tab:ionParameter},
e.g., the projected velocity
$-117\,$km\,s$^{-1}\le v_\mathrm{Ne\,X}\le -69\,$km\,s$^{-1}$
measured for \Ion{Ne}{x} is (during $0.93\le\phi_\mathrm{orb}\le1$)
obtained at $r_{12}=1.78\pm0.07$.
For many other ions, both results are still very similar:
e.g., \Ion{Ne}{ix} with $-214\,$km\,s$^{-1}\le v_\mathrm{Ne\,IX}\le -123\,$km\,s$^{-1}$
is expected at $r_{12}=1.95\pm0.07$ from the $v_\mathrm{rad,BH}$ model
and at $r_{12}=2.1$--2.8 from the photoionization model.
Only for the highly ionized iron lines, the small distances are
-- as already anticipated in \Sect{sec:XSTAR} --
underestimated by the \textsc{XSTAR} simulation run with constant average density,
which overestimates the wind density close to the black hole.
For example, the velocity range $-9\,$km\,s$^{-1}\le v_\mathrm{Fe\,XXIV}\le52\,$km\,s$^{-1}$
measured for \Ion{Fe}{xxiv} corresponds to $r_{12}=1.51\pm0.07$,
while the population of this ion peaks at $r_{12}=0.8$--1.1
within the model presented in Table~\ref{tab:ionParameter}.

\subsection{Previous \Chandra{} Observations}\label{sec:prevObs}
\begin{figure}\centering
 \includegraphics[width=0.95\columnwidth]{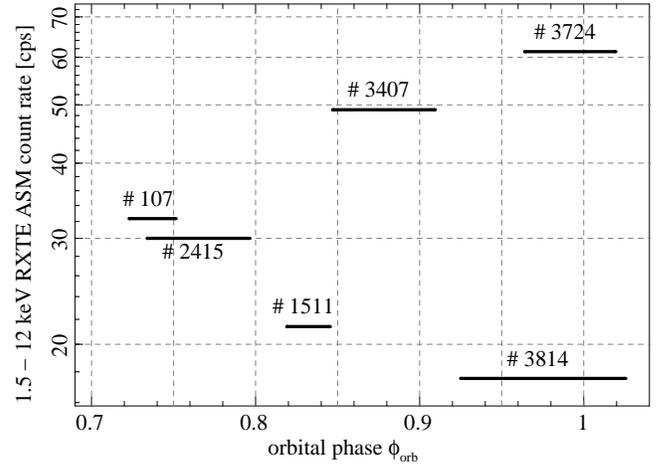}
 \caption{Orbital phase and mean ASM count rate
          of the previously reported \Chandra{} observations of \Cyg{}
          (ObsID~3814 is presented in this paper).
         }
 \label{fig:observations}
\end{figure}
In previous \Chandra-HETGS observations of \Cyg,
the stellar wind was seen under different viewing angles,
or the X-ray source was in other spectral states (see \Fig{fig:observations}),
which probably means that the properties of the wind were also different.
\citet{Schulz2002} have analyzed the 14\,ks \ifApJ{}{observation }ObsID~107
performed in 1999 October (at $\phi_\mathrm{orb}\approx0.74$),\footnote{%
 Note that \citet{Schulz2002} and \citet{Miller2005} quote an erroneous date
 and thus the wrong orbital phase $\phi_\mathrm{orb}=0.93\ne0.74$ for this observation.}
when \Cyg{} was in a transitional state.
They derive a neutral column density
of $N_\mathrm{H}=6.2\!\times\!10^{21}\:\mathrm{cm}^{-2}$
from prominent absorption edges (see also Table~\ref{tab:neutralAbundances})
and detect some emission and absorption lines with indications of P~Cygni profiles.
\citet{Miller2005} have investigated the focused wind
with the 32\,ks \ifApJ{}{observation }ObsID~2415 of \Cyg{} in an intermediate state,
performed in 2001 January (at $\phi_\mathrm{orb}\approx0.77$).
They report absorption and emission lines
of H- and He-like resonance lines of Ne, Na, Mg, and Si
with a mean redshift of $\approx$100\,km\,s$^{-1}$,
as well as some lines of highly ionized Fe and Ni.
The column density is $6.2\!\times\!10^{21}\:\mathrm{cm}^{-2}$
\citep{Miller2002}.
\citet{Marshall2001} describe Ly\,$\alpha$ and He\,$\alpha$ absorption lines,
redshifted by $(450\pm150)\:\mathrm{km\,s}^{-1}$
from the 12.6\,ks \ifApJ{}{observation }ObsID~1511 of \Cyg{} in the hard state,
performed in 2001 January (at $\phi_\mathrm{orb}\approx0.83$).
\citet{ChangCui2007} report dramatic variability
in the 30\,ks \ifApJ{}{observation }ObsID~3407 performed in 2001 October
(at $\phi_\mathrm{orb}\approx0.88$),
when \Cyg{} reached its soft state.
While a large number of absorption lines (mostly redshifted, but not by a consistent velocity)
is identified in the first part of their observation,
most of them weaken significantly or cannot be detected at all
during the second part.
The complete ionization of the wind due to a sudden density decrement
is given as a possible explanation.
\citet{Feng2003} detect asymmetric absorption lines
with the 26\,ks \ifApJ{}{observation }ObsID~3724 of \Cyg{} in the soft state,
performed in 2002 July (at $\phi_\mathrm{orb}=0$).
The line centers are almost at their rest wavelengths,
but the red wings are more extended,
especially for the transitions of highest ionized ions,
which they explain by the inflowing focused wind reaching
both the highest redshift and ionization parameter
closest to the black hole.

The interpretation of the absorption lines presented in the previous section
describes our observation consistent with wind and photoionization models.
It can, however, not be applied to all of the other observations;
the model of \Fig{fig:proj_velocity} neither predicts
a conspicuously high redshift at $\phi_\mathrm{orb}\approx0.83$ and at a higher soft X-ray flux
(ObsID~1511),
nor a positive redshift at $\phi_\mathrm{orb}\approx0.77$ at a still higher flux
(ObsID~2415, see \Fig{fig:observations}).
Inhomogeneities (e.g., density enhancements in shielding clumps)
and asymmetries of the wind (e.g., due to noninertial forces in the binary system)
may therefore play an important role in these cases.

\subsection{Summary}
In this paper, we have presented 
a \Chandra-HETGS observation of \Cyg{} during superior conjunction of the black hole,
which allows us to detect the X-ray absorption signatures of the stellar wind of HDE\,226868.
The light curve near $\phi_\mathrm{orb}=0$ is shaped by absorption dips;
these are, however, excluded from this analysis by selecting nondip times of high count rate only.
At a flux of $\sim$0.25\,Crab, we have to deal with moderate pile-up in the grating spectra,
for which can be accounted very well with the \texttt{simple\_gpile2} model.
The continuum of both \Chandra's soft and \RXTE's broadband X-ray spectrum
has been described consistently by a single model,
consisting of an empirical broken power-law spectrum with high-energy cutoff
(which is typical for the hard state)
and a subordinated disk component with an inner radius close to the ISCO.
The joint modeling of both spectra reveals the presence of both a narrow
and a relativistically broadened fluorescence Fe K$\alpha$ line.
\Chandra{} has resolved absorption edges of neutral atoms
with an overabundance of metals, suggesting an origin not only in the ISM,
but also in the stellar wind of the evolved supergiant.
The previously suspected anticorrelation of neutral column densities and X-ray flux,
which is due to photoionization, is confirmed.
Absorption lines of highly ionized ions are produced
where the wind becomes extremely photoionized.
For H- and He-like ions, Lyman and He-series are detected up to the Ly or He$\:\epsilon$ line,
which we use to measure the column density and velocity of absorbing ions.
For the wealth of Fe L-shell transitions,
column densities can best be obtained
by directly using a physical model for the complete line series of an ion.
The nondip spectrum shows almost no Doppler shifts,
probably indicating that we have excluded the focused wind by our selection of nondip times.
We have also detected two plasma components in emission
by two pairs of i- and f-lines of He-like Mg\,\textsc{xi}.
The first one is roughly at rest and
we identify it with the base of the spherical wind close to the stellar surface.
The second plasma component is denser and observed at a redshift
that is compatible with the focused wind.
A simple \textsc{XSTAR} simulation indicates that most of the observed ions
only exist in a distance of the black hole,
where the velocity of a spherical wind, projected against the black hole, is low
-- which is a consistent explanation of the small Doppler shifts observed in absorption lines.
We review the previously reported \Chandra{} observations of \Cyg{} and find
that not all of them can be described in the same picture,
as the wind may be affected by asymmetries or inhomogeneities.
A detailed spectroscopy of the absorption dips,
which might shed more light on the focused wind,
will be presented in a subsequent paper.

\acknowledgments
We are grateful to A.~Juett for her help with the data reduction
and her contributions to \verb|tbnew| model.
We thank A.~Young, J.~Xiang, and M.~B\"ock for helpful discussions,
J.~Houck for his support on \textsc{ISIS} and
T.~Kallman for his help with \textsc{XSTAR}.
M.\,H. and J.\,W. acknowledge funding
from the Bundesministerium f\"ur Wirtschaft und Technologie
through the Deutsches Zentrum f\"ur Luft- und Raumfahrt
under contract 50OR0701. M.\,N. is supported by NASA Grants GO3-4050B and SV3-73016.
We thank the International Space Science Institute, Berne, Switzerland,
and the MIT Kavli Institute for Space Research, Cambridge, MA, USA,
for their hospitality during the preparation of this work.



\begin{thebibliography}{}
\bibitem[\protect\astroncite{{Ba{\l}uci{\'n}ska-Church} et~al.}{1995}]{BalucinskaChurch1995}
 {Ba{\l}uci{\'n}ska-Church}, M., Belloni, T., Church, M.J., \& Hasinger, G. 1995,
 \href{http://adsabs.harvard.edu/abs/1995A\%26A...302L...5B}{A\&A, 302, L5}

\bibitem[\protect\astroncite{{Ba{\l}uci{\'n}ska-Church} et~al.}{2000}]{BalucinskaChurch2000}
 {Ba{\l}uci{\'n}ska-Church}, et~al. 2000,
 \href{http://adsabs.harvard.edu/abs/2000MNRAS.311..861B}{MNRAS, 311, 861}

\bibitem[\protect\astroncite{Belloni et~al.}{1996}]{Belloni1996}
 Belloni, T., et~al. 1996,
 \href{http://adsabs.harvard.edu/abs/1996ApJ...472L.107B}{ApJ, 472, L107}

\bibitem[\protect\astroncite{Blondin}{1994}]{Blondin1994}
 Blondin, J.M. 1994,
 \href{http://adsabs.harvard.edu/abs/1994ApJ...435..756B}{ApJ, 435, 756}

\bibitem[\protect\astroncite{Bolton}{1972}]{Bolton1972}
 Bolton, C.T. 1972,
 \href{http://adsabs.harvard.edu/abs/1972Natur.240..124B}{Nature, 240, 124}

\bibitem[\protect\astroncite{Bondi \& Hoyle}{1944}]{BondiHoyle1944}
 Bondi, H., \& Hoyle, F. 1944
 \href{http://adsabs.harvard.edu/abs/1944MNRAS.104..273B}{MNRAS, 104, 273}

\bibitem[\protect\astroncite{Bowyer et~al.}{1965}]{Bowyer1965}
 Bowyer, S., Byram, E.T., Chubb, T.A., \& {Friedman}, H. 1965,
 \href{http://adsabs.harvard.edu/abs/1965Sci...147..394B}{Science, 147, 394}

\bibitem[\protect\astroncite{Brocksopp et~al.}{1999}]{Brocksopp1999b}
 Brocksopp, et~al. 1999,
 \href{http://adsabs.harvard.edu/abs/1999MNRAS.309.1063B}{MNRAS, 309, 1063}

\bibitem[\protect\astroncite{{Cadolle Bel} et~al.}{2006}]{CadolleBel2006}
 {Cadolle Bel}, M., et~al. 2006,
 \href{http://adsabs.harvard.edu/abs/2006A\%26A...446..591C}{A\&A, 446, 591}

\bibitem[\protect\astroncite{Canizares et~al.}{2005}]{Canizares2005}
 Canizares, C.R., et~al. 2005,
 \href{http://adsabs.harvard.edu/abs/2005PASP..117.1144C}{PASP, 117, 1144}

\bibitem[\protect\astroncite{Castor et~al.}{1975}]{CastorAbbottKlein1975}
 Castor, J.I., Abbott, D.C., \& Klein, R.I. 1975,
 \href{http://adsabs.harvard.edu/abs/1975ApJ...195..157C}{ApJ, 195, 157}

\bibitem[\protect\astroncite{Chang \& Cui}{2007}]{ChangCui2007}
 Chang, C., \& Cui, W. 2007
 \href{http://adsabs.harvard.edu/abs/2007ApJ...663.1207C}{ApJ, 663, 1207}

\bibitem[\protect\astroncite{Conti}{1978}]{Conti1978}
 Conti, P.S. 1978
 \href{http://adsabs.harvard.edu/abs/1978A\%26A....63..225C}{A\&A, 63, 225}

\bibitem[\protect\astroncite{CXC}{2005}]{CXC2005_pileupABC}
 \Chandra{} X-ray Center 2005, The \Chandra{} ABC Guide to Pileup,
 \url{http://cxc.harvard.edu/ciao/download/doc/pileup\_abc.ps}

\bibitem[\protect\astroncite{CXC}{2006}]{CXC_POG}
 \Chandra{} X-ray Center 2006, The \Chandra{} Proposers' Observatory Guide,
 \url{http://cxc.harvard.edu/proposer/POG/}

\bibitem[\protect\astroncite{Davis}{2002}]{Davis2002}
 Davis, J.E. 2002,
 in High Resolution X-ray Spectroscopy with \textsl{XMM-Newton} and \Chandra,
 ed. G. {Branduardi-Raymont},
 \url{http://adsabs.harvard.edu/abs/2002hrxs.confE..11D}

\bibitem[\protect\astroncite{Davis}{2003}]{Davis2003}
 Davis, J.E. 2003,
 in \href{http://adsabs.harvard.edu/abs/2003SPIE.4851..101D}{Proc. SPIE 4851},
    X-Ray and Gamma-Ray Telescopes and Instruments for Astronomy,
 ed. J.E. Tr{\"u}mper, \& H.D. Tananbaum (Bellingham, WA:SPIE),
 \href{http://adsabs.harvard.edu/abs/2003SPIE.4851..101D}{101}

\bibitem[\protect\astroncite{Dotani et~al.}{1997}]{Dotani1997}
 Dotani, T., et~al. 1997
 \href{http://adsabs.harvard.edu/abs/1997ApJ...485L..87D}{ApJ, 485, L87}

\bibitem[\protect\astroncite{Doty}{1994}]{Doty1994}
 Doty, J.P. 1994,
 The All Sky Monitor for the X-ray Timing Explorer, Technical report, MIT,
 \url{http://xte.mit.edu/XTE.html}

\bibitem[\protect\astroncite{Feng et~al.}{2003}]{Feng2003}
 Feng, Y.X., Tennant, A.F., \& Zhang, S.N. 2003
 \href{http://adsabs.harvard.edu/abs/2003ApJ...597.1017F}{ApJ, 597, 1017}

\bibitem[\protect\astroncite{Friend \& Castor}{1982}]{FriendCastor1982}
 Friend, D.B., \& Castor, J.I. 1982
 \href{http://adsabs.harvard.edu/abs/1982ApJ...261..293F}{ApJ, 261, 293}

\bibitem[\protect\astroncite{Gabriel \& Jordan}{1969}]{GabrielJordan1969}
 Gabriel, A.H., \& Jordan, C. 1969,
 \href{http://adsabs.harvard.edu/abs/1969MNRAS.145..241G}{MNRAS, 145, 241}

\bibitem[\protect\astroncite{Garmire et~al.}{2003}]{Garmire2003}
 Garmire, G.P., et~al. 2003,
 in \href{http://adsabs.harvard.edu/abs/2003SPIE.4851...28G}{Proc. SPIE 4851},
    X-Ray and Gamma-Ray Telescopes and Instruments for Astronomy,
 ed. J.E. Truemper, \& H.D. Tananbaum (Bellingham, WA:SPIE),
 \href{http://adsabs.harvard.edu/abs/2003SPIE.4851...28G}{28}

\bibitem[\protect\astroncite{Gies \& Bolton}{1982}]{GiesBolton1982}
 Gies, D.R., \& Bolton, C.T. 1982
 \href{http://adsabs.harvard.edu/abs/1982ApJ...260..240G}{ApJ, 260, 240}

\bibitem[\protect\astroncite{Gies \& Bolton}{1986a}]{GiesBolton1986_II}
 Gies, D.R., \& Bolton, C.T. 1986a,
 \href{http://adsabs.harvard.edu/abs/1986ApJ...304..371G}{ApJ, 304, 371}

\bibitem[\protect\astroncite{Gies \& Bolton}{1986b}]{GiesBolton1986_III}
 Gies, D.R., \& Bolton, C.T. 1986b,
 \href{http://adsabs.harvard.edu/abs/1986ApJ...304..389G}{ApJ, 304, 389}

\bibitem[\protect\astroncite{Gies et~al.}{2003}]{Gies2003}
 Gies, D.R., et~al. 2003,
 \href{http://adsabs.harvard.edu/abs/2003ApJ...583..424G}{ApJ, 583, 424}

\bibitem[\protect\astroncite{Gies et~al.}{2008}]{Gies2008}
 Gies, D.R., et~al. 2008,
 \href{http://adsabs.harvard.edu/abs/2008ApJ...678.1237G}{ApJ, 678, 1237}

\bibitem[\protect\astroncite{Gleissner et~al.}{2004a}]{Gleissner2004_III}
 Gleissner, T., et~al. 2004a,
 \href{http://adsabs.harvard.edu/abs/2004A\%26A...425.1061G}{A\&A, 425, 1061}

\bibitem[\protect\astroncite{Gleissner et~al.}{2004b}]{Gleissner2004_II}
 Gleissner, T., et~al. 2004b,
 \href{http://adsabs.harvard.edu/abs/2004A\%26A...414.1091G}{A\&A 414, 1091}

\bibitem[\protect\astroncite{Gorczyca}{2000}]{Gorczyca2000}
 Gorczyca, T.W. 2000,
 \href{http://adsabs.harvard.edu/abs/2000PhRvA..61b4702G}{Phys. Rev. A, 61, 024702}

\bibitem[\protect\astroncite{Gorczyca \& McLaughlin}{2000}]{GorczycaMcLaughlin2000}
 Gorczyca, T.W., \& McLaughlin, B.M. 2000,
 \href{http://adsabs.harvard.edu/abs/2000JPhB...33L.859G}{J.~Phys.~B Atom.~Mol.~Phys.,}
 \href{http://adsabs.harvard.edu/abs/2000JPhB...33L.859G}{33, L859}

\bibitem[\protect\astroncite{Gruber et~al.}{1996}]{Gruber1996}
 Gruber, D.E., et~al. 1996,
 \href{http://adsabs.harvard.edu/abs/1996A\%26AS..120C.641G}{A\&AS, 120, C641}

\bibitem[\protect\astroncite{Hatchett \& McCray}{1977}]{HatchettMcCray1977}
 Hatchett, S., \& McCray, R. 1977,
 \href{http://adsabs.harvard.edu/abs/1977ApJ...211..552H}{ApJ, 211, 552}

\bibitem[\protect\astroncite{Herrero et~al.}{1995}]{Herrero1995}
 Herrero, A., et~al. 1995,
 \href{http://adsabs.harvard.edu/abs/1995A\%26A...297..556H}{A\&A, 297, 556}

\bibitem[\protect\astroncite{Houck}{2002}]{Houck2002}
 Houck, J.C. 2002,
 in High Resolution X-ray Spectroscopy with \textsl{XMM-Newton} and \Chandra,
 ed. G. {Branduardi-Raymont},
 \url{http://adsabs.harvard.edu/abs/2002hrxs.confE..17H}.

\bibitem[\protect\astroncite{Humphreys}{1978}]{Humphreys1978}
 Humphreys, R.M. 1978,
 \href{http://adsabs.harvard.edu/abs/1978ApJS...38..309H}{ApJS, 38, 309}

\bibitem[\protect\astroncite{Ishibashi}{2006}]{Bish2006}
 Ishibashi, K. 2006,
 Specification for finding zeroth order positions in grating data analysis,
 \url{http://space.mit.edu/CXC/docs/memo\_fzero\_1.3.ps}

\bibitem[\protect\astroncite{Jahoda et~al.}{1996}]{Jahoda1996}
 Jahoda, K., et~al. 1996,
 in \href{http://adsabs.harvard.edu/abs/1996SPIE.2808...59J}{Proc. SPIE 2808},
    EUV, X-Ray, and Gamma-Ray Instrumentation for Astronomy, Vol.~VII
 ed. O.H. Siegmund, \& M.A. Gummin (Bellingham, WA:SPIE),
 \href{http://adsabs.harvard.edu/abs/1996SPIE.2808...59J}{59}

\bibitem[\protect\astroncite{Juett et~al.}{2004}]{Juett2004}
 Juett, A.M., Schulz, N.S., \& Chakrabarty, D. 2004,
 \href{http://adsabs.harvard.edu/abs/2004ApJ...612..308J}{ApJ, 612, 308}

\bibitem[\protect\astroncite{Juett et~al.}{2006a}]{Juett2006}
 Juett, A.M., Schulz, N.S., Chakrabarty, D., \& Gorczyca, T.W. 2006a,
 \href{http://adsabs.harvard.edu/abs/2006ApJ...648.1066J}{ApJ, 648, 1066}

\bibitem[\protect\astroncite{Juett et~al.}{2006b}]{JuettWilms2006}
 Juett, A.M., Wilms, J., Schulz, N.S., \& Nowak, M.A. 2006b,
 \href{http://adsabs.harvard.edu/abs/2006AAS...209.1712J}{BAAS, 38, 921}

\bibitem[\protect\astroncite{Kahn et~al.}{2001}]{Kahn2001}
 Kahn, S.M., et~al. 2001,
 \href{http://adsabs.harvard.edu/abs/2001A\%26A...365L.312K}{A\&A, 365, L312}

\bibitem[\protect\astroncite{Kalberla et~al.}{2005}]{Kalberla2005}
 Kalberla, P.M.W., et~al. 2005,
 \href{http://adsabs.harvard.edu/abs/2005A\%26A...440..775K}{A\&A, 440, 775}

\bibitem[\protect\astroncite{Kallman \& Bautista}{2001}]{Kallman2001}
 Kallman, T., \& Bautista, M. 2001,
 \href{http://adsabs.harvard.edu/abs/2001ApJS..133..221K}{ApJS, 133, 221}

\bibitem[\protect\astroncite{Karitskaya et~al.}{2006}]{Karitskaya2006}
 Karitskaya, E.A., et~al. 2006,
 \href{http://adsabs.harvard.edu/abs/2006IBVS.5678....1K}{Inf. Bull. Variable Stars, 5678, 1}

\bibitem[\protect\astroncite{Kirsch et~al.}{2005}]{Kirsch2005}
 Kirsch, M.G., et~al. 2005,
 in \href{http://adsabs.harvard.edu/abs/2005SPIE.5898...22K}{Proc. SPIE 5898}
    UV, X-Ray, and Gamma-Ray Space Instrumentation for Astronomy, Vol.~XIV,
 ed. O.H.W. Siegmund (Bellingham, WA:SPIE),
 \href{http://adsabs.harvard.edu/abs/2005SPIE.5898...22K}{22}

\bibitem[\protect\astroncite{Kortright \& Kim}{2000}]{KortrightKim2000}
 Kortright, J.B., \& Kim, S.-K. 2000,
 \href{http://adsabs.harvard.edu/abs/2000PhRvB..6212216K}{Phys. Rev. B, 62, 12216}

\bibitem[\protect\astroncite{Kotani et~al.}{2000}]{Kotani2000}
 Kotani, T., et~al. 2000,
 \href{http://adsabs.harvard.edu/abs/2000ApJ...539..413K}{ApJ, 539, 413}

\bibitem[\protect\astroncite{Lachowicz et~al.}{2006}]{Lachowicz2006}
 Lachowicz, P., et~al. 2006,
 \href{http://adsabs.harvard.edu/abs/2006MNRAS.368.1025L}{MNRAS, 368, 1025}

\bibitem[\protect\astroncite{Lamers \& Leitherer}{1993}]{LamersLeitherer1993}
 Lamers, H.J.G.L.M., \& Leitherer, C. 1993,
 \href{http://adsabs.harvard.edu/abs/1993ApJ...412..771L}{ApJ, 412, 771}

\bibitem[\protect\astroncite{Lee et~al.}{2001}]{Lee2001}
 Lee, J.C., et~al. 2001,
 \href{http://adsabs.harvard.edu/abs/2001ApJ...554L..13L}{ApJ, 554, L13}

\bibitem[\protect\astroncite{Lee et~al.}{2002}]{Lee2002}
 Lee, J.C., et~al. 2002,
 \href{http://adsabs.harvard.edu/abs/2002ApJ...567.1102L}{ApJ, 567, 1102}

\bibitem[\protect\astroncite{{L{\'e}pine} \& Moffat}{2008}]{LepineMoffat2008}
 {L{\'e}pine}, S., \& Moffat, A.F.J. 2008,
 \href{http://adsabs.harvard.edu/abs/2008AJ....136..548L}{AJ, 136, 548}

\bibitem[\protect\astroncite{Levine et~al.}{1996}]{Levine1996}
 Levine, A.M., et~al. 1996,
 \href{http://adsabs.harvard.edu/abs/1996ApJ...469L..33L}{ApJ, 469, L33}

\bibitem[\protect\astroncite{Makishima et~al.}{2008}]{Makishima2008}
 Makishima, K., et~al. 2008,
 \href{http://adsabs.harvard.edu/abs/2008PASJ...60..585M}{PASJ, 60, 585}

\bibitem[\protect\astroncite{Marshall et~al.}{2001}]{Marshall2001}
 Marshall, H.L., Schulz, N.S., Fang, T., Cui, W., Canizares, C.R., Miller, J.M., \& Lewin, W.H.G. 2001,
 in \href{http://adsabs.harvard.edu/abs/2001xeab.confE..45M}{X-ray Emission from Accretion onto}
    \href{http://adsabs.harvard.edu/abs/2001xeab.confE..45M}{Black Holes},
 ed. T. Yaqoob, \& J.H. Krolik (Baltimore: John Hopkins Univ.),
 \href{http://adsabs.harvard.edu/abs/2001xeab.confE..45M}{45}

\bibitem[\protect\astroncite{Mauche et~al.}{2003}]{Mauche2003}
 Mauche, C.W., Liedahl, D.A., \& Fournier, K.B. 2003,
 \href{http://adsabs.harvard.edu/abs/2003ApJ...588L.101M}{ApJ, 588, L101}

\bibitem[\protect\astroncite{Mauche et~al.}{2005}]{Mauche2005}
 Mauche, C.W., Liedahl, D.A., \& Fournier, K.B. 2005,
 in \href{http://adsabs.harvard.edu/abs/2005AIPC..774..133M}{AIP Conf.~Ser. 774},
    X-ray Diagnostics of Astrophysical Plasmas: Theory, Experiment, and Observation,
 ed. R. Smith (New York: AIP),
 \href{http://adsabs.harvard.edu/abs/2005AIPC..774..133M}{133}

\bibitem[\protect\astroncite{McConnell et~al.}{2002}]{McConnell2002}
 McConnell, M.L., et~al. 2002,
 \href{http://adsabs.harvard.edu/abs/2002ApJ...572..984M}{ApJ, 572, 984}

\bibitem[\protect\astroncite{Merloni et~al.}{2000}]{MerloniFabian2000}
 Merloni, A., Fabian, A.C., \& Ross, R.R. 2000,
 \href{http://adsabs.harvard.edu/abs/2000MNRAS.313..193M}{MNRAS, 313, 193}

\bibitem[\protect\astroncite{Mewe \& Schrijver}{1978}]{MeweSchrijver1978}
 Mewe, R., \& Schrijver, J. 1978,
 \href{http://adsabs.harvard.edu/abs/1978A\%26A....65...99M}{A\&A, 65, 99}

\bibitem[\protect\astroncite{Mihalas}{1978}]{Mihalas_StellarAtmospheres}
 Mihalas, D. 1978,
 Stellar atmospheres (San Francisco, CA: W.H.~Freeman)

\bibitem[\protect\astroncite{Miller et~al.}{2002}]{Miller2002}
 Miller, J.M., et~al. 2002,
 \href{http://adsabs.harvard.edu/abs/2002ApJ...578..348M}{ApJ, 578, 348}

\bibitem[\protect\astroncite{Miller et~al.}{2005}]{Miller2005}
 Miller, J.M., et~al. 2005,
 \href{http://adsabs.harvard.edu/abs/2005ApJ...620..398M}{ApJ, 620, 398}

\bibitem[\protect\astroncite{Murdin \& Webster}{1971}]{MurdinWebster1971}
 Murdin, P., \& Webster, B.L. 1971,
 \href{http://adsabs.harvard.edu/abs/1971Natur.233..110M}{Nature, 233, 110}

\bibitem[\protect\astroncite{Ninkov et~al.}{1987a}]{Ninkov1987a}
 Ninkov, Z., Walker, G.A.H., \& Yang, S. 1987a,
 \href{http://adsabs.harvard.edu/abs/1987ApJ...321..425N}{ApJ, 321, 425}

\bibitem[\protect\astroncite{Ninkov et~al.}{1987b}]{Ninkov1987b}
 Ninkov, Z., Walker, G.A.H., \& Yang, S. 1987b
 \href{http://adsabs.harvard.edu/abs/1987ApJ...321..438N}{ApJ, 321, 438}

\bibitem[\protect\astroncite{Nowak et~al.}{1999}]{Nowak1999_II}
 Nowak, M.A., et~al. 1999,
 \href{http://adsabs.harvard.edu/abs/1999ApJ...510..874N}{ApJ, 510, 874}

\bibitem[\protect\astroncite{Nowak et~al.}{2008}]{Nowak2008}
 Nowak, M.A., et~al. 2008,
 \href{http://adsabs.harvard.edu/abs/2008ApJ...689.1199N}{ApJ, in press}
 (\href{http://adsabs.harvard.edu/abs/2008arXiv0809.3005N}{arXiv: 0809.3005})

\bibitem[\protect\astroncite{Oskinova et~al.}{2007}]{Oskinova2007}
 Oskinova, L.M., Hamann, W.-R., \& Feldmeier, A. 2007,
 \href{http://adsabs.harvard.edu/abs/2007A\%26A...476.1331O}{A\&A, 476, 1331}

\bibitem[\protect\astroncite{Porquet \& Dubau}{2000}]{PorquetDubau2000}
 Porquet, D., \& Dubau, J. 2000,
 \href{http://adsabs.harvard.edu/abs/2000A\%26AS..143..495P}{A\&AS, 143, 495}

\bibitem[\protect\astroncite{Pottschmidt et~al.}{2000}]{Pottschmidt2000}
 Pottschmidt, K., el~al. 2000,
 \href{http://adsabs.harvard.edu/abs/2000A\%26A...357L..17P}{A\&A, 357, L17}

\bibitem[\protect\astroncite{Pottschmidt et~al.}{2003}]{Pottschmidt2003}
 Pottschmidt, K., et~al. 2003,
 \href{http://adsabs.harvard.edu/abs/2003A%26A...407.1039P}{A\&A, 407, 1039}

\bibitem[\protect\astroncite{Poutanen et~al.}{2008}]{Poutanen2008}
 Poutanen, J., Zdziarski, A.A., \& Ibragimov, A. 2008,
 \href{http://adsabs.harvard.edu/abs/2008MNRAS.389.1427P}{MNRAS, 389, 1427}

\bibitem[\protect\astroncite{Remillard \& McClintock}{2006}]{RemillardMcClintock2006}
 Remillard, R.A., \& McClintock, J.E. 2006,
 \href{http://adsabs.harvard.edu/abs/2006ARA\%26A..44...49R}{ARA\&A, 44, 49}

\bibitem[\protect\astroncite{Russell et~al.}{2007}]{Russell2007}
 Russell, D.M., Fender, R.P., Gallo, E., \& Kaiser, C.R. 2007,
 \href{http://adsabs.harvard.edu/abs/2007MNRAS.376.1341R}{MNRAS, 376, 1341}

\bibitem[\protect\astroncite{Schulz et~al.}{2002}]{Schulz2002}
 Schulz, N.S., et al. 2002,
 \href{http://adsabs.harvard.edu/abs/2002ApJ...565.1141S}{ApJ, 565, 1141}

\bibitem[\protect\astroncite{Shaposhnikov \& Titarchuk}{2007}]{Shaposhnikov2007}
 Shaposhnikov, N., \& Titarchuk, L. 2007,
 \href{http://adsabs.harvard.edu/abs/2007ApJ...663..445S}{ApJ, 663, 445}

\bibitem[\protect\astroncite{Shimura \& Takahara}{1995}]{ShimuraTakahara1995}
 Shimura, T., \& Takahara, F. 1995,
 \href{http://adsabs.harvard.edu/abs/1995ApJ...445..780S}{ApJ, 445, 780}

\bibitem[\protect\astroncite{Spitzer}{1978}]{Spitzer1978_ISMprocesses}
 Spitzer, L. 1978,
 Physical processes in the interstellar medium (New York: Wiley-Interscience)

\bibitem[\protect\astroncite{{Str{\"o}mgren}}{1939}]{Stromgren1939}
 {Str{\"o}mgren}, B. 1939,
 \href{http://cdsads.u-strasbg.fr/abs/1939ApJ....89..526S}{ApJ, 89, 526}

\bibitem[\protect\astroncite{Szostek \& Zdziarski}{2007}]{Szostek2007}
 Szostek, A., \& Zdziarski, A.A. 2007,
 \href{http://adsabs.harvard.edu/abs/2007MNRAS.375..793S}{MNRAS, 375, 793}

\bibitem[\protect\astroncite{Treves et~al.}{1980}]{Treves1980}
 Treves, A., et~al. 1980,
 \href{http://adsabs.harvard.edu/abs/1980ApJ...242.1114T}{ApJ, 242, 1114}

\bibitem[\protect\astroncite{{van Aken} \& Liebscher}{2002}]{vanAkenLiebscher2002}
 van Aken, P.A., \& Liebscher, B. 2002,
 \href{http://adsabs.harvard.edu/abs/2002PCM....29..188V}{Phys. Chem. Miner., 29, 188}

\bibitem[\protect\astroncite{Verner et~al.}{1996}]{Verner1996}
 Verner, D.A., Verner, E.M., \& Ferland, G.J. 1996,
 \href{http://adsabs.harvard.edu/abs/1996ADNDT..64....1V}{At. Data Nucl. Data Tables, 64, 1}

\bibitem[\protect\astroncite{Verner \& Yakovlev}{1995}]{VernerYakovlev1995}
 Verner, D.A., \& Yakovlev, D.G. 1995,
 \href{http://adsabs.harvard.edu/abs/1995A\%26AS..109..125V}{A\&AS, 109, 125}

\bibitem[\protect\astroncite{Verner et~al.}{1993}]{Verner1993}
 Verner, D.A., Yakovlev, D.G., Band, I.M., \& Trzhaskovskaya, M.B. 1993,
 \href{http://adsabs.harvard.edu/abs/1993ADNDT..55....233V}{At. Data Nucl. Data Tables, 55, 233}

\bibitem[\protect\astroncite{Vrtilek et~al.}{2008}]{Vrtilek2008}
 Vrtilek, S.D., et~al. 2008,
 \href{http://adsabs.harvard.edu/abs/2008ApJ...678.1248V}{ApJ, 678, 1248}

\bibitem[\protect\astroncite{Walborn}{1973}]{Walborn1973}
 Walborn, N.R. 1973,
 \href{http://adsabs.harvard.edu/abs/1973ApJ...179L.123W}{ApJ, 179, L123}

\bibitem[\protect\astroncite{Webster \& Murdin}{1972}]{WebsterMurdin1972}
 Webster, B.L., \& Murdin, P. 1972,
 \href{http://adsabs.harvard.edu/abs/1972Natur.235...37W}{Nature, 235, 37}

\bibitem[\protect\astroncite{Wilms et~al.}{2000}]{Wilms2000}
 Wilms, J., Allen, A., \& McCray, R. 2000,
 \href{http://adsabs.harvard.edu/abs/2000ApJ...542..914W}{ApJ, 542, 914}

\bibitem[\protect\astroncite{Wilms et~al.}{2006}]{Wilms2006}
 Wilms, J., et~al. 2006,
 \href{http://adsabs.harvard.edu/abs/2006A%26A...447..245W}{A\&A, 447, 245}

\bibitem[\protect\astroncite{Wojdowski et~al.}{2008}]{Wojdowski2008}
 Wojdowski, P.S., Liedahl, D.A., \& {Kallman}, T.R. 2008,
 \href{http://adsabs.harvard.edu/abs/2008ApJ...673.1023W}{ApJ, 673, 1023}

\bibitem[\protect\astroncite{Xiang et~al.}{2005}]{Xiang2005}
 Xiang, J., Zhang, S.N., \& Yao, Y. 2005,
 \href{http://adsabs.harvard.edu/abs/2005ApJ...628..769X}{ApJ, 628, 769}

\bibitem[\protect\astroncite{Zhang et~al.}{1997}]{Zhang1997}
 Zhang, S.N., et~al. 1997,
 \href{http://adsabs.harvard.edu/abs/1997ApJ...477L..95Z}{ApJ, 477, L95}

\bibitem[\protect\astroncite{{Zi{\'o}{\l}kowski}}{2005}]{Ziolkowski2005}
 {Zi{\'o}{\l}kowski}, J. 2005,
 \href{http://adsabs.harvard.edu/abs/2005MNRAS.358..851Z}{MNRAS, 358, 851}
\end{thebibliography}
\end{document}